\documentclass[preprint2]{aastex}
\slugcomment{March 17 2015}

\usepackage{hyperref}
\usepackage{amsmath}
\usepackage{graphicx}
\usepackage{datetime}
\usepackage{amssymb}
\usepackage{color}

\newcommand{\aS}{\texttt{autoScan}}

\newcommand{\snana}{\texttt{SNANA}}
\newcommand{\q}[1]{\texttt{#1}}

\newcommand{\med}{\mathrm{med}}

\newcommand{\about}{{\raise.17ex\hbox{$\scriptstyle\sim$}}}

\begin{document}
\slugcomment{\today}
\title{Automated Transient Identification in the Dark Energy Survey}
\author{
D.~A.~Goldstein,\altaffilmark{1,2}
C.~B.~D'Andrea,\altaffilmark{3}
J.~A.~Fischer,\altaffilmark{4}
R.~J.~Foley,\altaffilmark{5,6}
R.~R.~Gupta,\altaffilmark{7}
R.~Kessler,\altaffilmark{8,9}
A.~G.~Kim,\altaffilmark{2}
R.~C.~Nichol,\altaffilmark{3}
P.~Nugent,\altaffilmark{1,2}
A.~Papadopoulos,\altaffilmark{3}
M.~Sako,\altaffilmark{4}
M.~Smith,\altaffilmark{10}
M.~Sullivan,\altaffilmark{10}
R.~C.~Thomas,\altaffilmark{2}
W.~Wester,\altaffilmark{11}
R.C.~Wolf,\altaffilmark{4}
F.~B.~Abdalla,\altaffilmark{12}
M.~Banerji,\altaffilmark{13,14}
A.~Benoit-L{\'e}vy,\altaffilmark{12}
E.~Bertin,\altaffilmark{15}
D.~Brooks,\altaffilmark{12}
A.~Carnero~Rosell,\altaffilmark{16,17}
F.~J.~Castander,\altaffilmark{18}
L.~N.~da Costa,\altaffilmark{16,17}
R.~Covarrubias,\altaffilmark{19}
D.~L.~DePoy,\altaffilmark{20}
S.~Desai,\altaffilmark{21}
H.~T.~Diehl,\altaffilmark{11}
P.~Doel,\altaffilmark{12}
T.~F.~Eifler,\altaffilmark{4,22}
A.~Fausti Neto,\altaffilmark{16}
D.~A.~Finley,\altaffilmark{11}
B.~Flaugher,\altaffilmark{11}
P.~Fosalba,\altaffilmark{18}
J.~Frieman,\altaffilmark{8,11}
D.~Gerdes,\altaffilmark{23}
D.~Gruen,\altaffilmark{24,25}
R.~A.~Gruendl,\altaffilmark{5,19}
D.~James,\altaffilmark{26}
K.~Kuehn,\altaffilmark{27}
N.~Kuropatkin,\altaffilmark{11}
O.~Lahav,\altaffilmark{12}
T.~S.~Li,\altaffilmark{20}
M.~A.~G.~Maia,\altaffilmark{16,17}
M.~Makler,\altaffilmark{28}
M.~March,\altaffilmark{4}
J.~L.~Marshall,\altaffilmark{20}
P.~Martini,\altaffilmark{29,30}
K.~W.~Merritt,\altaffilmark{11}
R.~Miquel,\altaffilmark{31,32}
B.~Nord,\altaffilmark{11}
R.~Ogando,\altaffilmark{16,17}
A.~A.~Plazas,\altaffilmark{22,33}
A.~K.~Romer,\altaffilmark{34}
A.~Roodman,\altaffilmark{35,36}
E.~Sanchez,\altaffilmark{37}
V.~Scarpine,\altaffilmark{11}
M.~Schubnell,\altaffilmark{23}
I.~Sevilla-Noarbe,\altaffilmark{5,37}
R.~C.~Smith,\altaffilmark{26}
M.~Soares-Santos,\altaffilmark{11}
F.~Sobreira,\altaffilmark{11,16}
E.~Suchyta,\altaffilmark{29,38}
M.~E.~C.~Swanson,\altaffilmark{19}
G.~Tarle,\altaffilmark{23}
J.~Thaler,\altaffilmark{6}
A.~R.~Walker\altaffilmark{26}
}
\altaffiltext{1}{Department of Astronomy, University of California, Berkeley, 501 Campbell Hall \#3411, Berkeley, CA 94720}
\altaffiltext{2}{Lawrence Berkeley National Laboratory, 1 Cyclotron Road, Berkeley, CA 94720, USA}
\altaffiltext{3}{Institute of Cosmology and Gravitation, University of Portsmouth, Dennis Sciama Building, Burnaby Road, Portsmouth, PO1 3FX, UK}
\altaffiltext{4}{Department of Physics and Astronomy, University of Pennsylvania, Philadelphia, PA 19104, USA}
\altaffiltext{5}{Astronomy Department, University of Illinois at Urbana-Champaign, 1002 W.\ Green Street, Urbana, IL 61801, USA}
\altaffiltext{6}{Department of Physics, University of Illinois at Urbana-Champaign, 1110 W.\ Green Street, Urbana, IL 61801, USA}
\altaffiltext{7}{Argonne National Laboratory, 9700 South Cass Avenue, Lemont, IL 60439, USA}
\altaffiltext{8}{Kavli Institute for Cosmological Physics, University of Chicago, Chicago, IL 60637, USA}
\altaffiltext{9}{Department of Astronomy and Astrophysics, University of Chicago, 5640 South Ellis Avenue, Chicago, IL 60637, USA}
\altaffiltext{10}{School of Physics and Astronomy, University of Southampton, Highfield, Southampton, SO17 1BJ, UK }
\altaffiltext{11}{Fermi National Accelerator Laboratory, P. O. Box 500, Batavia, IL 60510, USA}
\altaffiltext{12}{Department of Physics \& Astronomy, University College London, Gower Street, London, WC1E 6BT, UK}
\altaffiltext{13}{Kavli Institute for Cosmology, University of Cambridge, Madingley Road, Cambridge CB3 0HA, UK}
\altaffiltext{14}{Institute of Astronomy, University of Cambridge, Madingley Road, Cambridge CB3 0HA, UK}
\altaffiltext{15}{Institut d'Astrophysique de Paris, Univ. Pierre et Marie Curie \& CNRS UMR7095, F-75014 Paris, France}
\altaffiltext{16}{Laborat\'orio Interinstitucional de e-Astronomia - LIneA, Rua Gal. Jos\'e Cristino 77, Rio de Janeiro, RJ - 20921-400, Brazil}
\altaffiltext{17}{Observat\'orio Nacional, Rua Gal. Jos\'e Cristino 77, Rio de Janeiro, RJ - 20921-400, Brazil}
\altaffiltext{18}{Institut de Ci\`encies de l'Espai, IEEC-CSIC, Campus UAB, Facultat de Ci\`encies, Torre C5 par-2, 08193 Bellaterra, Barcelona, Spain}
\altaffiltext{19}{National Center for Supercomputing Applications, 1205 West Clark St., Urbana, IL 61801, USA}
\altaffiltext{20}{George P. and Cynthia Woods Mitchell Institute for Fundamental Physics and Astronomy, and Department of Physics and Astronomy, Texas A\&M University, College Station, TX 77843,  USA}
\altaffiltext{21}{Department of Physics, Ludwig-Maximilians-Universitaet, Scheinerstr. 1, 81679 Muenchen, Germany}
\altaffiltext{22}{Jet Propulsion Laboratory, California Institute of Technology, 4800 Oak Grove Dr., Pasadena, CA 91109, USA}
\altaffiltext{23}{Department of Physics, University of Michigan, Ann Arbor, MI 48109, USA}
\altaffiltext{24}{Max Planck Institute for Extraterrestrial Physics, Giessenbachstrasse, 85748 Garching, Germany}
\altaffiltext{25}{University Observatory Munich, Scheinerstrasse 1, 81679 Munich, Germany}
\altaffiltext{26}{Cerro Tololo Inter-American Observatory,
National Optical Astronomy Observatory,
Casilla 603, Colina El Pino S/N, La Serena, Chile}
\altaffiltext{27}{Australian Astronomical Observatory, North Ryde, NSW 2113, Australia}
\altaffiltext{28}{ICRA, Centro Brasileiro de Pesquisas F\'isicas, Rua Dr. Xavier Sigaud 150, CEP 22290-180, Rio de Janeiro, RJ, Brazil}
\altaffiltext{29}{Center for Cosmology and Astro-Particle Physics, The Ohio State University, Columbus, OH 43210, USA}
\altaffiltext{30}{Department of Astronomy, The Ohio State University, Columbus, OH 43210, USA}
\altaffiltext{31}{Institut de F\'{\i}sica d'Altes Energies, Universitat Aut\`onoma de Barcelona, E-08193 Bellaterra, Barcelona, Spain}
\altaffiltext{32}{Instituci\'o Catalana de Recerca i Estudis Avan\c{c}ats, E-08010 Barcelona, Spain}
\altaffiltext{33}{Brookhaven National Laboratory, Bldg 510, Upton, NY 11973, USA}
\altaffiltext{34}{Astronomy Centre, University of Sussex, Falmer, Brighton, BN1 9QH, UK}
\altaffiltext{35}{Kavli Institute for Particle Astrophysics \& Cosmology, P. O. Box 2450, Stanford University, Stanford, CA 94305, USA}
\altaffiltext{36}{SLAC National Accelerator Laboratory, Menlo Park, CA 94025, USA}
\altaffiltext{37}{Centro de Investigaciones Energ\'eticas, Medioambientales y Tecnol\'ogicas (CIEMAT), Madrid, Spain}
\altaffiltext{38}{Department of Physics, The Ohio State University, Columbus, OH 43210, USA}
\begin{abstract}
We describe an algorithm for identifying point-source transients and 
moving objects on reference-subtracted optical
images containing artifacts of processing and instrumentation. The
algorithm makes use of the supervised machine learning technique
known as Random Forest. We present results from its use in the Dark Energy
Survey Supernova program (DES-SN), where it was trained using a sample of 898,963 signal and
background events generated by the transient detection pipeline. After
reprocessing the data collected during the
first DES-SN observing season (Sep. 2013 through Feb. 2014) using the
algorithm, the number of transient candidates eligible for human scanning decreased
by a factor of 13.4, while only 1.0 percent of the artificial Type Ia supernovae (SNe)
injected into search images to monitor survey efficiency 
were lost, most of which were very faint events. 
Here we characterize the
algorithm's performance in detail, and we discuss
how it can inform pipeline design decisions for future
time-domain imaging surveys, such as the Large Synoptic Survey
Telescope and the Zwicky Transient Facility. An implementation of the algorithm 
and the training data used in this paper are available at \color{blue}{
\href{http://portal.nersc.gov/project/dessn/autoscan}{http://portal.nersc.gov/project/dessn/autoscan}}.
\end{abstract}


\section{Introduction}
\label{section:intro}
To identify scientifically valuable transients or moving objects on the sky, imaging
surveys have historically adopted a manual approach, employing humans to visually inspect images for signatures of the
events (e.g., \citealt{zwicky, calan+tololo, scpposter, highz, kait, higherz, blanc, snls, sakosdss, neowise, comets, panstarrssn}). 
But recent advances in the capabilities of
telescopes, detectors, and supercomputers have 
fueled a dramatic rise
in the data production rates of
such surveys, straining
the ability of their teams to quickly and comprehensively look at
images to perform discovery.

\begin{figure*}
\centering
\includegraphics[width=160mm]{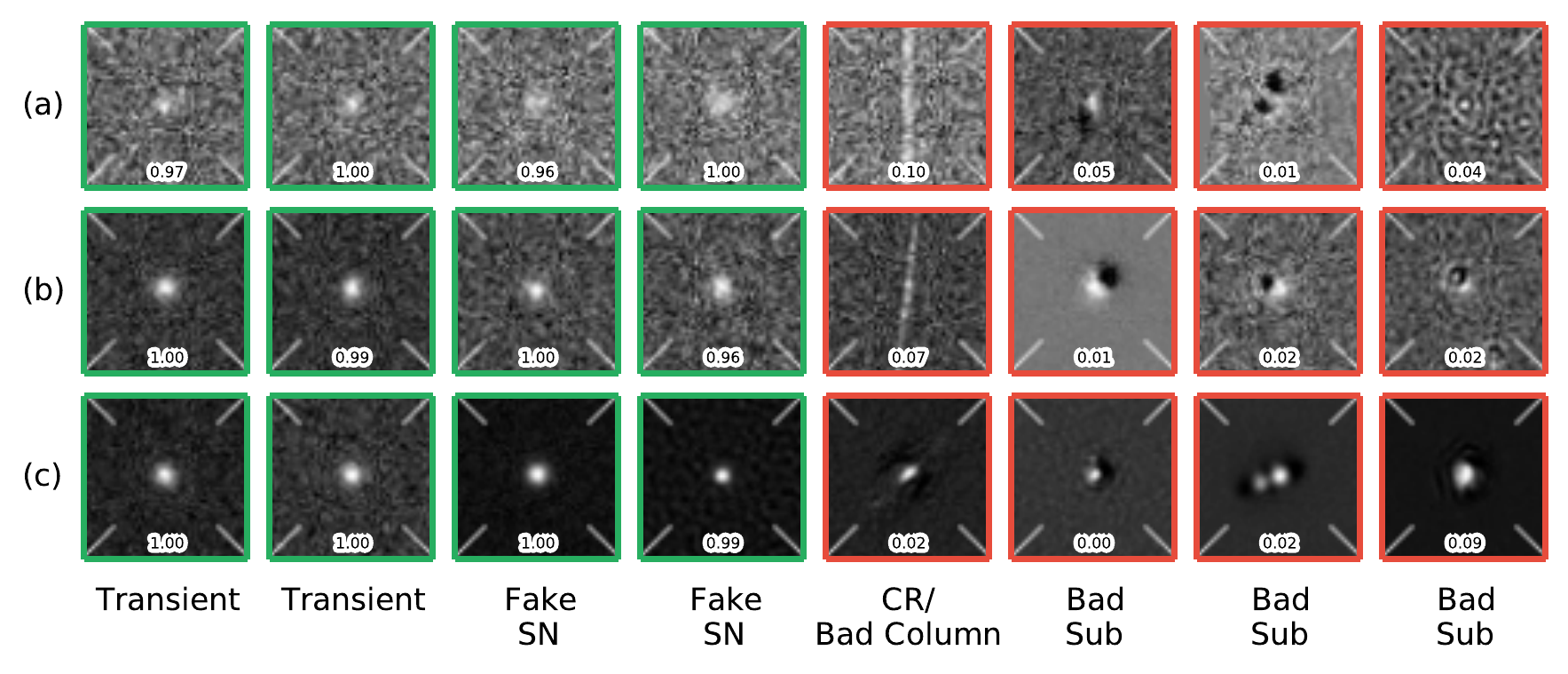}
\caption{Cutouts  of  DES  difference
  images, roughly 14 arcsec on a side, centered on legitimate (green
  boxes; left four columns of figure) and spurious (red boxes; right
  four columns of figure) objects, at a
  variety of signal-to-noise ratios: (a) $\mathrm{S/N} \leq 10$, (b)
  $10 < \mathrm{S/N} \leq 30$, (c) $30 < \mathrm{S/N} \leq 100$.  The
  cutouts are subclassed to illustrate both the visual diversity of
  spurious objects and the homogeneity of
  authentic ones. Objects in the ``Transient''
  columns are real astrophysical transients
  that subtracted cleanly.
  Objects in the ``Fake SN'' columns are
  fake SNe Ia injected
  into transient search images to monitor
  survey efficiency.
  The column labeled ``CR/Bad Column'' 
  shows detections of cosmic rays (rows
  b and c) and a bad column on the
  CCD detector (row a). The
  columns labeled ``Bad Sub'' show non-varying astrophysical
  sources that did not subtract cleanly; this can result from poor astrometric
  solutions, shallow templates, or 
  bad observing conditions. 
  The numbers at the bottom of each cutout indicate
  the score that each detection received from the machine learning
  algorithm introduced in \S\ref{section:autoscan}; a score of 1.0
  indicates the algorithm is perfectly confident that
  the detection is not
  an artifact, while a score of 0.0 indicates the opposite.}
\label{fig:objs}
\end{figure*}

For surveys that search for objects on difference images---CCD images that
reveal changes in the appearance of a region of the sky between two
points in time---this problem of data volume is compounded by the
problem of data purity. Difference images are produced by subtracting reference images from single-epoch images
in a
process that involves
point-spread function (PSF) matching and image distortion (see,
e.g., \citealt{diffim}). In addition to legitimate detections of astrophysical variability, they can
contain artifacts of the differencing process, such as poorly
subtracted galaxies, and artifacts of the single-epoch images, such
as cosmic rays, optical ghosts, star halos, defective
pixels, near-field objects, and CCD edge effects.
Some examples are presented in Figure \ref{fig:objs}.
These artifacts can
vastly outnumber the signatures of scientifically valuable sources
on the images,
forcing object detection thresholds to be considerably higher than what
is to be expected from Gaussian fluctuations. 

For time-domain imaging surveys with a spectroscopic follow-up
program, these issues of data volume and purity
are compounded by time-pressure to
produce lists of the most promising targets for follow-up observations before they
become too faint to observe or fall outside a window of scientific
utility.  Ongoing searches for Type Ia supernovae (SNe Ia) out to
$z\sim1$, such as the Panoramic Survey Telescope and Rapid Response
System Medium Deep Survey's \citep{panstarrssn} and the Dark Energy Survey's
(DES; \citealt{des}), face all three of these challenges.
The DES supernova program (DES-SN; \citealt{bernsteinetal}), for example, produces up to 170 gigabytes 
of raw imaging data on a nightly basis. Visual
examination of sources extracted from the resulting
difference images using \q{SExtractor}
\citep{sextractor} revealed that $\about93$ percent are artifacts, even
after selection cuts (Kessler et al. 2015, in preparation). 
Additionally, the survey has a science-critical spectroscopic follow-up
program for
which it must routinely
select the $\about10$ most promising transient candidates from hundreds of
possibilities, most of which are artifacts.
This program is crucial to
survey science as it allows DES to confirm transient candidates as SNe,
train and optimize its photometric SN typing algorithms
(e.g., \q{PSNID}; \citealt{psnid}, \q{NNN}; \citealt{karpenka}),
and investigate interesting non-SN transients. 
To prepare a list of objects eligible for consideration for spectroscopic follow-up observations, members of DES-SN 
scanned 
nearly 1 million objects extracted from difference images during the survey's first observing season,
the numerical equivalent of 
nearly a week of uninterrupted scanning time, 
assuming scanning one object takes half a second.

For DES to meet its discovery goals, more efficient
techniques for artifact rejection on difference images are
needed.
Efforts to ``crowd-source'' similar large-scale classification
problems have been successful at scaling with growing data rates;
websites such as \url{Zooniverse.org} have accumulated over one
million users to tackle a variety of astrophysical classification
problems, including the classification of transient candidates from the
Palomar Transient Factory (PTF; \citealt{zoo}).  However, for DES to
optimize classification accuracy and generate reproducible 
classification decisions, automated
techniques are required.  

To reduce
the number of spurious candidates considered for spectroscopic follow-up, many
surveys impose selection requirements on quantities that can be
directly and automatically
computed from the raw imaging data.  Making hard selection
cuts of this kind has been shown to be a suboptimal technique for
artifact rejection in difference imaging. Although such cuts are
automatic and easy to interpret, they do not naturally handle
correlations between features, and they are an inefficient way to
select a subset of the high-dimensional feature space as the number of
dimensions grows large \citep{bailey}.

In contrast to selection cuts, machine learning (ML) classification
techniques provide a flexible solution to the problem of artifact
rejection in difference imaging. In general, these techniques attempt
to infer a precise mapping between numeric features that describe
characteristics of observed data, and the classes or labels assigned to those data, using a training set of feature-class pairs. ML
classification algorithms that generate decision rules using labeled
data---data whose class membership has already been
definitively established---are called ``supervised'' algorithms.
After generating a decision rule, supervised ML classifiers can be
used to predict the classes of unlabeled data instances. For a
review of supervised ML classification in astronomy, see,
e.g. \cite{astroml}. For an
introduction to the statistical underpinnings of supervised ML
classification techniques, see \cite{mlbook}.

Such classifiers address many of the shortcomings of scanning and
selection cuts.  ML algorithms' decisions are automatic, reproducible,
and fast enough to process streaming data in real-time. Their biases
can be systematically and quantitatively studied, and, most
importantly, given adequate computing resources, they remain fast and
consistent in the face of increasing data production rates. As more
data are collected, ML methods can continue to refine their knowledge
about a data set (see \S \ref{section:future}), thereby improving their
predictive performance on future data.  Supervised ML classification
techniques are currently used in a variety of astronomical contexts,
including time-series analysis, such as the
classification of variable stars \citep{varstars} and
SNe \citep{karpenka} from light curves, and
image analysis,
such as the typing of galaxies \citep{banerji}, and discovery of
trans-Neptunian objects (Gerdes et al. 2015, in preparation) on images. Although their
input data types differ, light curve shape  and
image-based ML classification frameworks are quite
similar: both operate on tabular numeric classification features
computed from raw input data (see \S\ref{section:features}). 

The use of supervised machine learning classification techniques for artifact
rejection in difference imaging was pioneered
by \cite{bailey} for the Nearby Supernova Factory \citep{snfactory}
using imaging data from the Near-Earth Asteroid Tracking
program\footnote{\url{http://neat.jpl.nasa.gov}.} and the
Palomar-QUEST Consortium, using the 112-CCD QUEST-II
camera \citep{baltay07}.  They compared the performance of three
supervised classification techniques---a Support Vector Machine, a
Random Forest, and an ensemble of boosted decision trees---in
separating a combination of real and fake detections of SNe from
background events.  They found that boosted decision trees constructed
from a library of astrophysical domain features (magnitude, FWHM,
distance to the nearest object in the reference co-add, measures of
roundness, etc.)  provided the best overall performance.

\cite{bloom12} built on the methodology of \cite{bailey} 
 by developing a highly accurate Random Forest framework for
classifying detections of variability extracted from PTF difference
images. \cite{brink} made
improvements to the classifier of \cite{bloom12}, setting an unbroken
benchmark for best overall performance on the PTF data set, using the
technique of recursive feature elimination to optimize their
classifier.  Recently, \cite{pca} published a systematic comparison of
several classification algorithms using features based on Principal
Component Analysis (PCA) extracted from Sloan Digital Sky Survey-II SN
survey difference images. Finally,
\cite{wright} used a pixel-based approach
to engineer a Random Forest classifier
for the Pan-STARRS Medium Deep Survey.

In this article, we describe \aS, a computer program developed for this purpose in DES-SN.
Our main objective is to report
the methodology that DES-SN adopted to construct an effective supervised
classifier, with an eye toward informing the design of similar
frameworks for future time domain surveys such as the Large Synoptic
Survey Telescope (LSST; \citealt{lsstsci}) and the Zwicky Transient Facility 
(ZTF; \citealt{ztf}).  We
extend the work of previous authors to a newer, larger data set,
showing how greater selection efficiency can be achieved by
increasing training set size, using generative models
for training data, and implementing new classification
features.

The structure of the paper is as follows. In \S \ref{section:des}, we provide an
overview of DES and the DES-SN transient detection
pipeline. In \S \ref{section:autoscan}, 
we describe the development
of \aS.
In \S \ref{section:performance}, we present metrics for evaluating the code's
performance and review its performance on a realistic classification
task. In \S \ref{section:discussion}, we discuss 
lessons learned and areas of future development that
can inform the design of similar frameworks for future surveys.

\section{The Dark Energy Survey and Transient Detection Pipeline}
\label{section:des}
In this section, we introduce DES
and the DES-SN transient detection pipeline (``{\tt DiffImg}''; Kessler et al. 2015, in preparation), which produced
the data used to train and validate \aS.
DES is a Stage III ground-based dark energy experiment designed to
provide the tightest constraints to date on the dark energy equation
of state parameter using observations of the four most
powerful probes of dark energy suggested by the Dark Energy Task Force
(DETF; \citealt{DETF}): SNe~Ia, galaxy clusters, baryon acoustic
oscillations, and weak gravitational lensing.  DES consists of two
interleaved imaging surveys: a wide-area survey that covers 5,000
deg$^2$ of the south Galactic cap in 5 filters $(grizY)$, and DES-SN, a
time-domain transient survey that covers 10 (8 ``shallow'' and 2 ``deep'') 3
deg$^2$ fields in the XMM-LSS, ELAIS-S, CDFS, and Stripe-82 regions of the sky, in four filters $(griz)$.  
The survey's main instrument,
the Dark Energy Camera (DECam; \citealt{diehl, flaugherdecam}; Flaugher et al. 2015, submitted), is a 570-megapixel 3
deg$^2$ imager with 62 fully depleted, red-sensitive CCDs.  It is
mounted at the prime focus of the Victor M. Blanco 4m telescope at the Cerro
Tololo Inter-American Observatory (CTIO).  DES conducted ``science
verification'' (SV) commissioning observations from November 2012
until February 2013, and it began science operations in August 2013
that will continue until at least 2018 \citep{diehly1}. The data used in this 
article are from the first season of DES science operations (``Y1"; 
Aug. 2013---Feb. 2014).

\begin{figure*}
\centering
\includegraphics{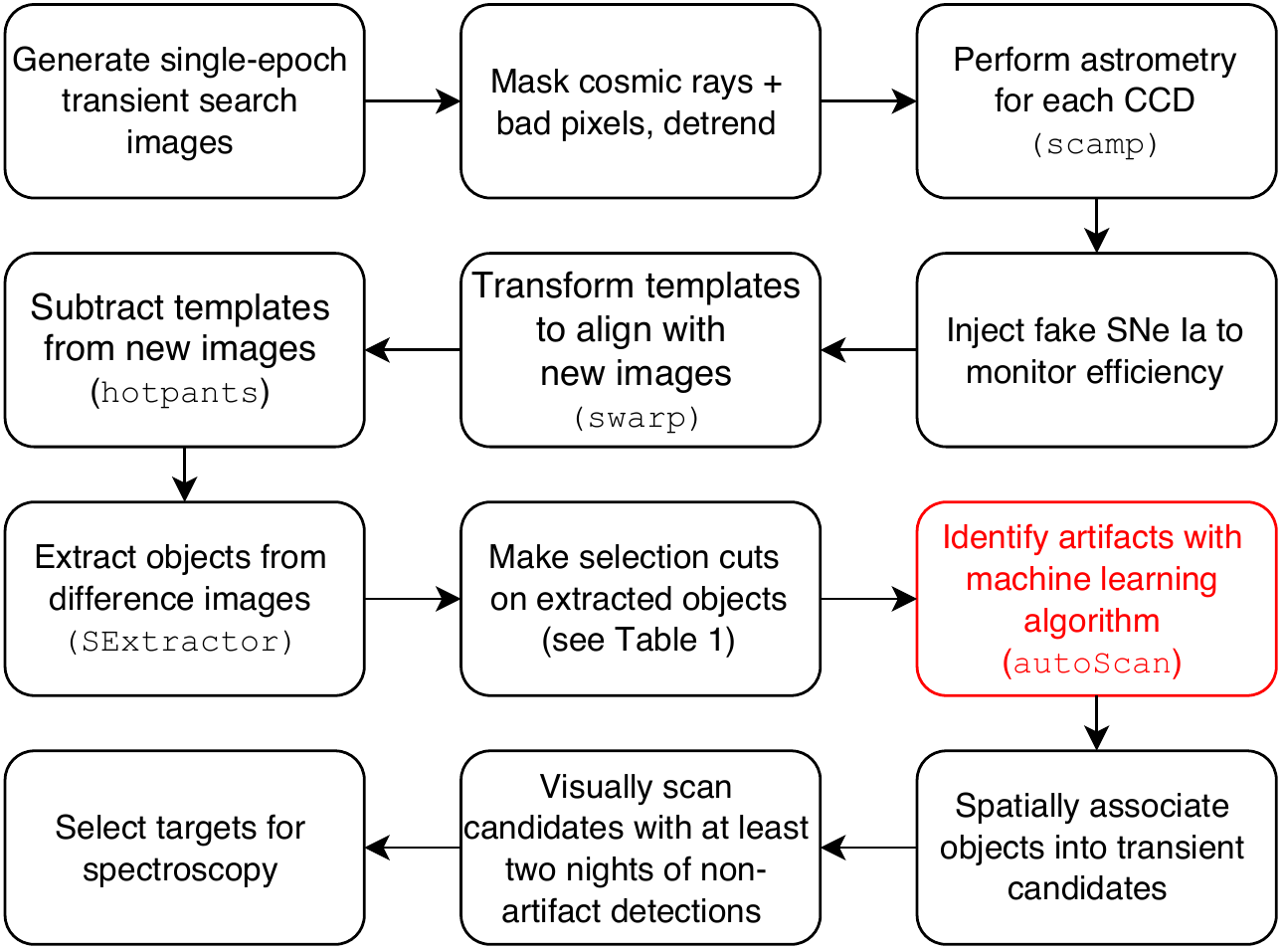}
\caption{Schematic of the DES-SN transient
detection pipeline.
         The magnitudes of fake SNe~Ia used to
         monitor survey efficiency 
         are calibrated using the zero point
         of the images into which they are injected
         and generated according to the procedure
         described in \S \ref{section:trainingdata}.  The \aS~step
         (red box) occurs after selection cuts are applied to objects
         extracted from difference images and before objects are
         spatially associated into raw
         transient candidates. Codes used
         at specific steps are indicated in
         parenthesis.}
\label{fig:diffim}
\end{figure*}

A schematic of the pipeline that DES-SN employs to discover transients 
is presented in Figure \ref{fig:diffim}. Transient survey 
``science images'' are single-epoch CCD images from the DES-SN fields.
After the image subtraction
step, sources are extracted using \q{SExtractor}.  Sources that pass
the cuts described in the
Object section of Table \ref{tab:cuts} are
referred to as ``detections." 
A ``raw candidate" is defined when
two or more detections match to within 
1''. A raw candidate
is promoted to a ``science candidate'' 
when it passes the \texttt{NUMEPOCHS}
requirement in Table \ref{tab:cuts}.
This
selection requirement was imposed to reject Solar System objects, such
as main belt asteroids and Kuiper belt objects, which move
substantially on images from night to night. 
Science candidates are eligible for visual 
examination and spectroscopic follow-up observations.
During the observing season, science candidates
are routinely photometered, fit with multi-band SN
light curve models, visually inspected, and slated for spectroscopic follow-up.

\begin{deluxetable}{llp{20mm}lp{67mm}}
\tablecaption{DES-SN object and candidate selection 
		      requirements.\label{tab:cuts}}
\tablehead{\colhead{Set} & \colhead{Feature} & \colhead{Lower Limit} & \colhead{Upper Limit} & \colhead{Description} }
\startdata
  Object & 
  \q{MAG} & \nodata & 30.0 & Magnitude from \q{SExtractor}. \\
  &\q{A\_IMAGE} & \nodata & 1.5 pix. & Length of semi-major axis from \q{SExtractor}. \\
  &\q{SPREAD\_MODEL} & \nodata & $3\sigma_S + 1.0$ & Star-galaxy separation output parameter from \q{SExtractor}.
                                                $\sigma_S$ is the estimated \q{SPREAD\_MODEL} uncertainty.\\
  &\q{CHISQ} & \nodata & $10^4$ &
            $\chi^2$ from PSF-fit to $35 \times 35$ pixel cutout around object in difference image. \\
  &\q{SNR}           & 3.5 & \nodata & Flux from a PSF-model fit to a $35 \times 35$ pixel 
  cutout around the object divided by the uncertainty 
  from the fit.\\
  &\q{VETOMAG}\tablenotemark{a}
  & 21.0 & \nodata & Magnitude from \texttt{SExtractor} for use in veto catalog check. \\
  &\q{VETOTOL}\tablenotemark{a} & Magnitude-dependent & \nodata & Separation from nearest object in veto catalog of bright stars.\\
  & \q{DIPOLE6} & \nodata & 2 & $N_{pix}$ in $35\times35$ pixel object-centered cutout at least 6$\sigma$ below 0.\\
  & \q{DIPOLE4} & \nodata & 20 & $N_{pix}$ in $35\times35$ pixel object-centered cutout at least 4$\sigma$ below 0.\\
  & \q{DIPOLE2} & \nodata & 200 & $N_{pix}$ in $35 \times 35$ pixel object-centered cutout at least 2$\sigma$ below 0.\\
  Candidate &
  \q{NUMEPOCHS} & 2 & \nodata & Number of distinct nights that the candidate is detected.
\enddata
\tablenotetext{a}{The difference imaging pipeline is expected to produce
								   false positives
                                   near bright or variable stars, thus all difference image objects
                                   are checked against a ``veto'' catalog of known
                                   bright and variable stars and are rejected if they are brighter than 21st magnitude and within a magnitude-dependent radius of a veto catalog source. 
                                   Thus only one of \q{VETOMAG} and \q{VETOTOL} 
   				 				   must be satisfied for an object to be
                 				   selected.}
\end{deluxetable}

\section{Classifier Development}
\label{section:autoscan}
In this section, we describe the
development of \aS.
We present
the classifier's training data set
(\S\ref{section:trainingdata}),
its classification feature set (\S\ref{section:featandproc}), and
the selection (\S\ref{section:algorithms}), properties (\S\ref{section:importances}), and optimization (\S\ref{section:optimization}) of its core
classification algorithm.

\subsection{Training Data}
\label{section:trainingdata}

To make probabilistic statements about the class membership
of new data, supervised ML classifiers must be trained or fit to
existing data whose true class labels are already known.
Each data instance is described by numeric classification
``features" (see \S\ref{section:features}); an effective
training data set must approximate the joint feature distributions of all
classes considered. Objects
extracted from difference images can belong to one
of two classes: ``Artifacts," or ``Non-Artifacts." 
Examples of each class must be present in the 
training set. 
Failing to include data from
certain regions of feature space can corrode the predictive
performance of the classifier in those regions, introducing bias into
the search that can systematically degrade survey
efficiency \citep{active}. Because the training set compilation
described here took place during the beginning of
Y1, it was complicated by
a lack of available visually scanned ``non-artifact" sources.

Fortunately, labeling data does not necessarily require humans to
visually inspect images. \cite{bloom12} discuss a variety of methods
for labeling detections of variability produced by difference imaging
pipelines, including scanning alternatives such as artificial source
construction and spectroscopic follow-up. Scanning, spectroscopy, and using
fake data each have their respective merits and drawbacks. Scanning is
laborious and potentially inaccurate, especially if each data instance
is only examined by one scanner, or if scanners are not well
trained. However, a large group of scanners can quickly label a number
of detections sufficient to create a training set for a machine
classifier, and \cite{brink} have shown that the supervised
classification algorithm Random Forest, which was ultimately selected
for \aS, is insensitive to mislabeled training data up to a
contamination level of 10 percent.

Photometric typing (e.g., \citealt{psnid}) can also be
useful for labeling detections of transients. However, robust
photometric typing requires well-sampled light curves, which in turn
require high-cadence photometry of difference image objects over
timescales of weeks or months. This requirement is prohibitive for
imaging surveys in their early stages. Further, because photometric
typing is an integral part of the spectroscopic target selection
process, by extension new imaging surveys also have too few detections
of spectroscopically confirmed SNe, AGN, or variable
stars. Native spectroscopic training samples are therefore impractical
sources of training data for new surveys.

Artificial source construction is the fastest method for generating native
detections of non-artifact sources in the early stages of a
survey. Large numbers of artificial transients (``fakes'')
can be injected into survey science images, and by construction 
their associated detections are true positives. Difficulties can arise when
the joint feature distributions of fakes selected for the training set
do not approximate the joint feature distributions of observed
transients in production. In DES-SN, SN~Ia fluxes from fake SN~Ia
light curves are overlaid on images near real galaxies. The fake SN~Ia
light curves are generated by the \snana~simulation \citep{snana},  and they
include true parent populations of stretch and color, a realistic
model of intrinsic scatter, a redshift range from 0.1 to 1.4, and a 
galaxy location proportional to surface brightness.
On difference images, detections of overlaid fakes
are visually indistinguishable from real point-source
transients and Solar System objects moving slowly enough
not to streak.
All fake SN~Ia
light curves are generated and stored prior to the start of the
survey. The overlay procedure is part of the difference imaging
pipeline, where the SN~Ia flux added to the image is scaled by the
zero point, spread over nearby pixels using a model of the PSF, and fluctuated by
random Poisson noise. These fakes are used to monitor the single-epoch transient
detection efficiency, as well as the candidate efficiency in which
detections on two distinct nights are required. On average, six detections
of fake SNe are overlaid on each single-epoch CCD-image.

The final \aS~training set contained detections of visually scanned
artifacts and artificial sources only. We did not include detections
of photometrically typed transients to minimize the contamination of
the ``Non-Artifact'' class with false positives. \cite{bailey} also used a
training set in which the ``Non-Artifact'' class consisted largely of artificial
sources. 

With 898,963
training instances in total, the \aS~training set is the largest used
for difference image artifact rejection in production. It was split
roughly evenly between ``real'' and ``artifact'' labeled
instances---454,092 were simulated SNe Ia injected onto host galaxies,
while the remaining 444,871 detections were human-scanned
artifacts. Compiling a set of artifacts to train \aS~was accomplished
by taking a random sample of the objects that had been scanned as
artifacts by humans during an early processing of DES Y1 data 
with a pared-down version of the difference imaging pipeline presented
in Figure \ref{fig:diffim}.

\subsection{Features and Processing}
\label{section:featandproc}
The supervised learning algorithms we consider in this analysis are
nonlinear functions that map points representing
individual detections in feature space to points in a space of object classes or class probabilities. The second design choice
in developing \aS~is therefore to define a suitable feature space
in which to represent the data instances we wish to use for training,
validation, and prediction. In this section, we describe the classification 
features that
we computed from the raw output of the difference imaging pipeline, 
as well as the steps used to pre- and post-process these features.

\subsubsection{Data Preprocessing}
\label{section:preprocessing}
The primary data sources for \aS~features are $51 \times 51$ pixel
object-centered search, template, and difference image cutouts. The template and difference image cutouts
are sky-subtracted. The search image cutout is sky-subtracted
if and only if it does not originate from a coadded exposure, though
this is irrelevant for what follows as no features
are directly computed from search image pixel values. 
Photometric measurements, \q{SExtractor} output parameters, and
other data sources are also used. Each
cutout associated with a detection is compressed to $25 \times
25$ pixels. The seeing for each search image is usually no less than 1
arcsec, while the DECam pixel scale lies between 0.262 and 0.264
arcsec depending on the location on the focal plane, so little
information is lost during compression. Although some artifacts are sharper than the seeing, we found that using compressed cutouts to compute some features resulted in better performance. 

Consider
a search, template, or difference
image cutout
associated with a single detection. 
Let the matrix element $\mathbf{I}_{x,y}$
of the $51 \times 51$ matrix $\mathbf{I}$
 represent the flux-value of the pixel at location $x, y$ on the cutout. We adopt the convention
 of zero-based indexing and the convention
 that element (0, 0) corresponds to the pixel at the top
 left-hand corner of the cutout.
Let the matrix element $C_{x,y}$ of the $25 \times 25$ matrix $C$ represent the flux-value of the pixel at location $x,y$ on the compressed cutout. Then $C$ is defined element-wise from $\mathbf{I}$ via
\begin{equation}
        C_{x,y} = \frac{1}{N_u} \sum_{i=0}^1 \sum_{j=0}^1 \mathbf{I}_{2x
        + i, 2y + j},
\label{eq:compress}        
\end{equation}
where $N_u$ is the number of unmasked pixels in the sum. 
Masked pixels are excluded from the sum.  Only when all four terms in the sum represent masked pixels is the
corresponding pixel masked in $C$. Note that matrix elements from the
right-hand column and last row of $\mathbf{I}$ never
appear in Equation \ref{eq:compress}.

To ensure that the pixel flux-values across cutouts are comparable, we
rescale the pixel values of each compressed cutout via
"';
\begin{equation}
        R_{x,y} = \frac{C_{x,y} - \med(C)}{\hat{\sigma}},
\end{equation} 
where the matrix element $R_{x,y}$ of the $25 \times 25$ matrix $R$
represents the flux-value of the pixel at location $x,y$ on the compressed, rescaled
cutout, and $\hat{\sigma}$ is a consistent estimator of the standard
deviation of $C$.  We take the median absolute
deviation as a consistent estimator of the standard deviation \citep{madpaper},
according to
\begin{equation}
        \hat{\sigma} = \frac{\med(|C-\med(C)|)}{\Phi^{-1}\left(\frac{3}{4}\right)}
\end{equation}
where $1/\Phi^{-1}(3/4) \approx 1.4826$ is the reciprocal of the inverse cumulative distribution for the
standard normal distribution evaluated at $3/4$. This is done to ensure that the effects
of defective pixels and cosmic rays nearly perpendicular to the focal plane are suppressed.  We therefore
have the following closed-form expression for the matrix element
$R_{x,y}$,
\begin{equation}
        R_{x, y} \approx \frac{1}{1.4826}\left[\frac{C_{x,y}
          - \med(C)}{\med(|C - \med(C)|)}\right] .
\label{eq:rescale}
\end{equation}
The rescaling expresses the value of each pixel on the compressed cutout as the
number of standard deviations above the median. Masked pixels are 
excluded from the computation of the
median in Equation \ref{eq:rescale}.

Finally, an additional rescaling from \cite{brink} is defined according to
\begin{equation}
        \mathbf{B}_{x, y} = \frac{\mathbf{I}_{x, y}
        - \med(\mathbf{I})}{\max(|\mathbf{I}|)}
\label{eq:rescale2}
\end{equation}
The size of $\mathbf{B}$ is $51 \times 51$. 
We found that using $\mathbf{B}$ instead of $R$ or $\mathbf{I}$ to
compute certain features resulted in better classifier performance.
Masked pixels are 
excluded from the computation of the median in
Equation \ref{eq:rescale2}.

\subsubsection{Feature Library}
\label{section:features}
Two feature libraries were investigated. The first was
primarily ``pixel-based.'' For a given object,  each matrix element of the rescaled, compressed
search, template, and difference cutouts 
was used as a
feature. The CCD ID number of each detection was also used, as DECam has 62 CCDs with specific artifacts (such as bad columns and hot pixels) as well as effects that are reproducible on the same CCD depending on which field is observed (such as bright stars). The
signal-to-noise ratio of each detection was also used as a feature.
The merits of this feature space include relatively straightforward
implementation and computational efficiency. A production version of
this pixel-based classifier was implemented in the DES-SN transient
detection pipeline at the beginning of Y1. In production, it became
apparent that the 1,877-dimensional\footnote{625 pixels on a $25 
\times 25$ pixel cutout $\times$ 3 cutouts per detection + 2 non-pixel features (\q{snr}, \q{ccdid}) = 1,877.} feature space was dominated by
uninformative features, and that better false positive control could be
achieved with a more compact feature set.

We pursued an alternative feature space going forward, instead using
38 high-level metrics to characterize detections of variability. 
A subset of the features are based on analogs
from \cite{bloom12} and \cite{brink}. In this section, we describe the
features that are new.  We present an at-a-glance view of the
entire \aS~feature library in Table \ref{tab:features}.  Histograms
  and contours for the three most important features
in the final \aS~model (see \S\ref{section:importances}) appear in
Figure \ref{fig:top3features}.

\begin{deluxetable}{lllp{90mm}}
\tablecaption{\q{autoScan}'s feature library. \label{tab:features}}
\tablehead{\colhead{Feature Name} & 
		   \colhead{Importance} & 
           \colhead{Source} & 
           \colhead{Description}}
\startdata           
\q{r\_aper\_psf} & 0.148 & New & The average flux in a 5-pixel
circular aperture centered on the object on the $\mathbf{I}^t$ cutout plus the flux from a $35\times35$-pixel PSF model-fit to the object on the $\mathbf{I}^d$ cutout, all divided by the 
PSF model-fit flux.  \\
\q{magdiff} & 0.094 & B12 & If a
source is found within 5'' of the location of the object in the galaxy coadd catalog, the
difference between \q{mag} and the magnitude of the nearby
source. Else, the difference between \q{mag} and the limiting
magnitude of the parent image from which the $\mathbf{I}^{d}$
cutout was generated.
\\
 \q{spread\_model} & 0.066 & New & \q{SPREAD\_MODEL} output
parameter from \q{SExtractor} on $\mathbf{I}^{d}$. \\
 \q{n2sig5} &
0.055 & B12 & Number of matrix elements in a $7 \times 7$ element block centered on the
detection on $R^{d}$ with values less than -2. \\
\q{n3sig5} & 0.053 & B12 &
Number of matrix elements in a $7 \times 7$ element block centered on the detection on
$R^{d}$ with values less than -3. \\
 \q{n2sig3} & 0.047
& B12 & Number of matrix elements in a $5 \times 5$ element block centered on
the detection on $R^{d}$ with values less than -2. \\
 \q{flux\_ratio} & 0.037 & B12 &  Ratio of the flux in a 5-pixel circular
aperture centered on the location of the detection on
$\mathbf{I}^{d}$ to the absolute value of the flux in a 5-pixel circular at the
same location on $\mathbf{I}^{t}$.  \\ 
\q{n3sig3} &
0.034 &  B12 & Number of matrix elements in a $5 \times 5$ element block centered on the
detection on $R^{d}$ with values less than -3. \\
 \q{mag\_ref\_err} & 0.030 & B12
& Uncertainty on \q{mag\_ref}, if it exists. Else imputed.  \\
\q{snr} & 0.029 & B12 & The flux from a $35 \times 35$-pixel 
PSF model-fit to the object on $\mathbf{I}^d$ divided by the
uncertainty from the fit. \\
 \q{colmeds} &
0.028 & New &  The maximum of the median pixel values of each column
on $\mathbf{B}^{d}$.  \\
 \q{nn\_dist\_renorm} & 0.027 & B12 & The distance from the
detection to the nearest source in the galaxy coadd catalog, if one
exists within 5''.  Else imputed. \\
\q{ellipticity} & 0.027 & B12 &
The ellipticity of the detection on $\mathbf{I}^{d}$ using \q{a\_image}~and
\q{b\_image} from \q{SExtractor}. \\
 \q{amp} & 0.027 & B13 & Amplitude of fit
that produced \q{gauss}. \\
 \q{scale} & 0.024 & B13 & Scale parameter of fit
that produced \q{gauss}. \\
 \q{b\_image}
& 0.024 & B12 &  Semi-minor axis of object from \q{SExtractor} on $\mathbf{I}^{d}$.  \\
 \q{mag\_ref} & 0.022 & B12 & The magnitude of
the nearest source in the galaxy coadd catalog, if one exists within
5'' of the detection on $\mathbf{I}^{d}$.  Else imputed.
\\
 \q{diffsum}
& 0.021 & New & The sum of the matrix elements in a $5 \times 5$ element
box centered on the detection location on $R^{d}$. \\
 \q{mag} & 0.020 & B12 & The magnitude of
the object from \q{SExtractor} on $\mathbf{I}^d$. \\
 \q{a\_ref} &
0.019 & B12 & Semi-major axis of 
the nearest source in the galaxy coadd catalog, if one
exists within 5''.  Else imputed.\tablebreak \\
 \q{n3sig3shift} & 0.019 & New &
The number of matrix elements with values greater than or equal to
$3$ in the central $5 \times 5$ element block of $R^d$ minus the
number of matrix elements with values greater than or equal to $3$
in the central $5 \times 5$ element block of $R^t$. \\
 \q{n3sig5shift} & 0.018 & New
& The number of matrix elements with values greater than or equal to
$3$ in the central $7 \times 7$ element block of $R^d$ minus the
number of matrix elements with values greater than or equal to $3$
in the central $7 \times 7$ element block of $R^t$
\\
\q{n2sig3shift} & 0.014 & New &
The number of matrix elements with values greater than or equal to
$2$ in the central $5 \times 5$ element block of $R^d$ minus the
number of matrix elements with values greater than or equal to $2$
in the central $5 \times 5$ element block of $R^t$. \\
  \q{b\_ref} &
0.012 &  B12 & Semi-minor axis of the 
nearest source in the galaxy coadd catalog, if one
exists within 5''.  Else imputed. \\
 \q{gauss} & 0.012 &  B13 & $\chi^2$ from fitting a
spherical, 2D Gaussian to a $15 \times 15$ pixel cutout around
the detection on $\mathbf{B}^{d}$. \\
 \q{n2sig5shift} & 0.012 & New &
The number of matrix elements with values greater than or equal to
$2$ in the central $7 \times 7$ element block of $R^d$ minus the
number of matrix elements with values greater than or equal to $2$
in the central $7 \times 7$ element block of $R^t$. \\
 \q{mag\_from\_limit} & 0.010 & B12 &  Limiting magnitude of the
parent image from which the $I^{d}$ cutout was generated minus \q{mag}. \\
 \q{a\_image} & 0.009 &  B12 & Semi-major axis of
object on $\mathbf{I}^{d}$ from \q{SExtractor}.
\\
 \q{min\_dist\_to\_edge} & 0.009 & B12 & Distance in pixels to the
nearest edge of the detector array on the parent image from which the $I^{d}$ cutout was generated.
\\
 \q{ccdid} & 0.008 & B13 & The numerical ID of the CCD on which
the detection was registered.  \\
 \q{flags} & 0.008 & B12 &
Numerical representation of \q{SExtractor} extraction flags on
$\mathbf{I}^{d}$. \\
 \q{numneg} & 0.007 & New &  The number of
negative matrix elements in a $7\times7$ element box centered on the detection
in $R^{d}$. \\
\q{l1} & 0.006 & B13 & $\mathrm{sign}(\sum \mathbf{B}^d) \times \sum |\mathbf{B}^d| / |\sum \mathbf{B}^d|$\\
 \q{lacosmic} & 0.006 & New &  $\max(\mathbf{B}^{d}) /
\max(F)$, where $F$ is the \q{LACosmic} \citep{lacosmic} ``fine
structure'' image computed on $\mathbf{B}^{d}$.
\\
 \q{spreaderr\_model} & 0.006 & New &  Uncertainty on \q{spread\_model}.
\\
 \q{maglim} & 0.005 & B12 & True if there is no nearby galaxy
coadd source, false otherwise. \\
 \q{bandnum} & 0.004 &  New &
Numerical representation of image filter. \\
 \q{maskfrac} & 0.003 &
New & The fraction of $\mathbf{I}^{d}$ that is
masked. \\
\enddata
\tablecomments{Source column indicates the
  reference in which the feature was first published. B13 indicates
  the feature first appeared in \protect\cite{brink}; B12 indicates the
  feature first appeared in \protect\cite{bloom12}, and New indicates the
  feature is new in this work. See \S\ref{section:algorithms} for an
  explanation of how feature importances are computed. Imputation refers
  to the procedure described in \S\ref{section:postprocessing}.}
\end{deluxetable}

\subsubsection{New Features}
\begin{figure}
\centering
\plotone{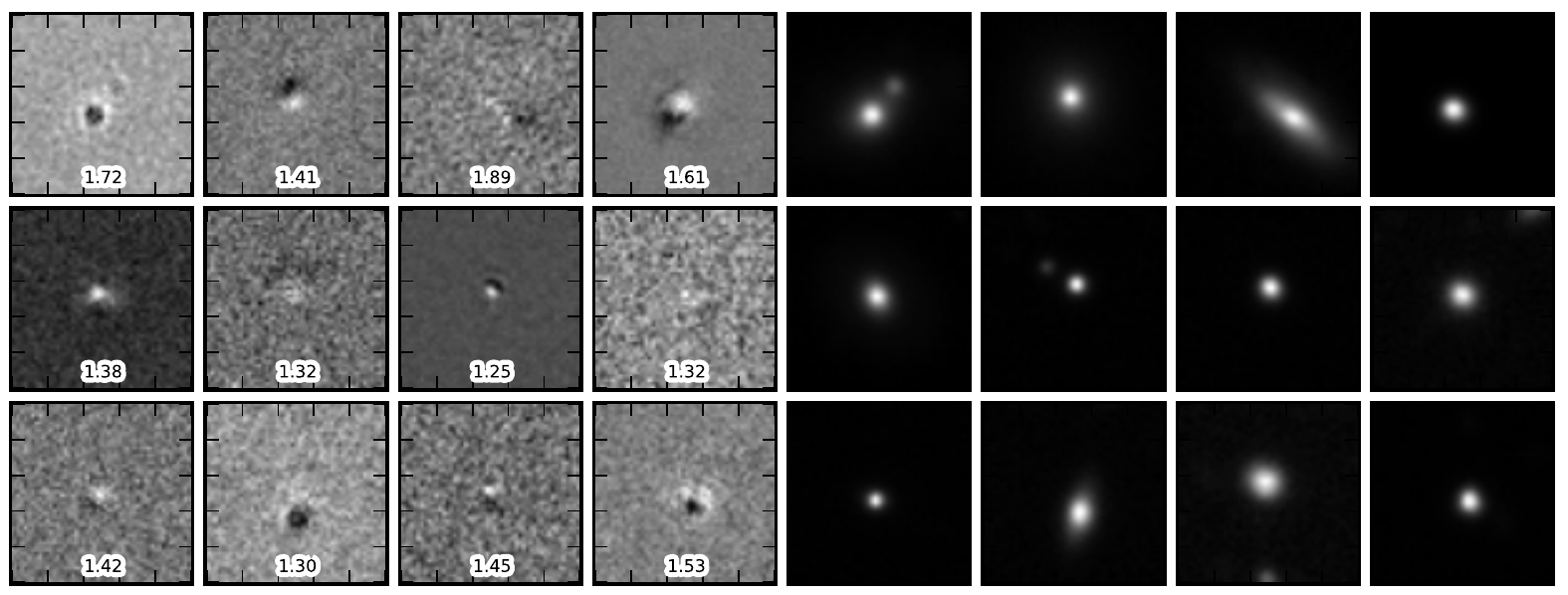}
\caption{Difference image cutouts (left four columns;
\q{r\_aper\_psf} values indicated) and 
corresponding template
image cutouts (right four columns) for objects with \q{r\_aper\_psf}
$> 1.25$.}
\label{fig:raperpsf}
\end{figure}

\begin{figure*}
 \centering
 \plotone{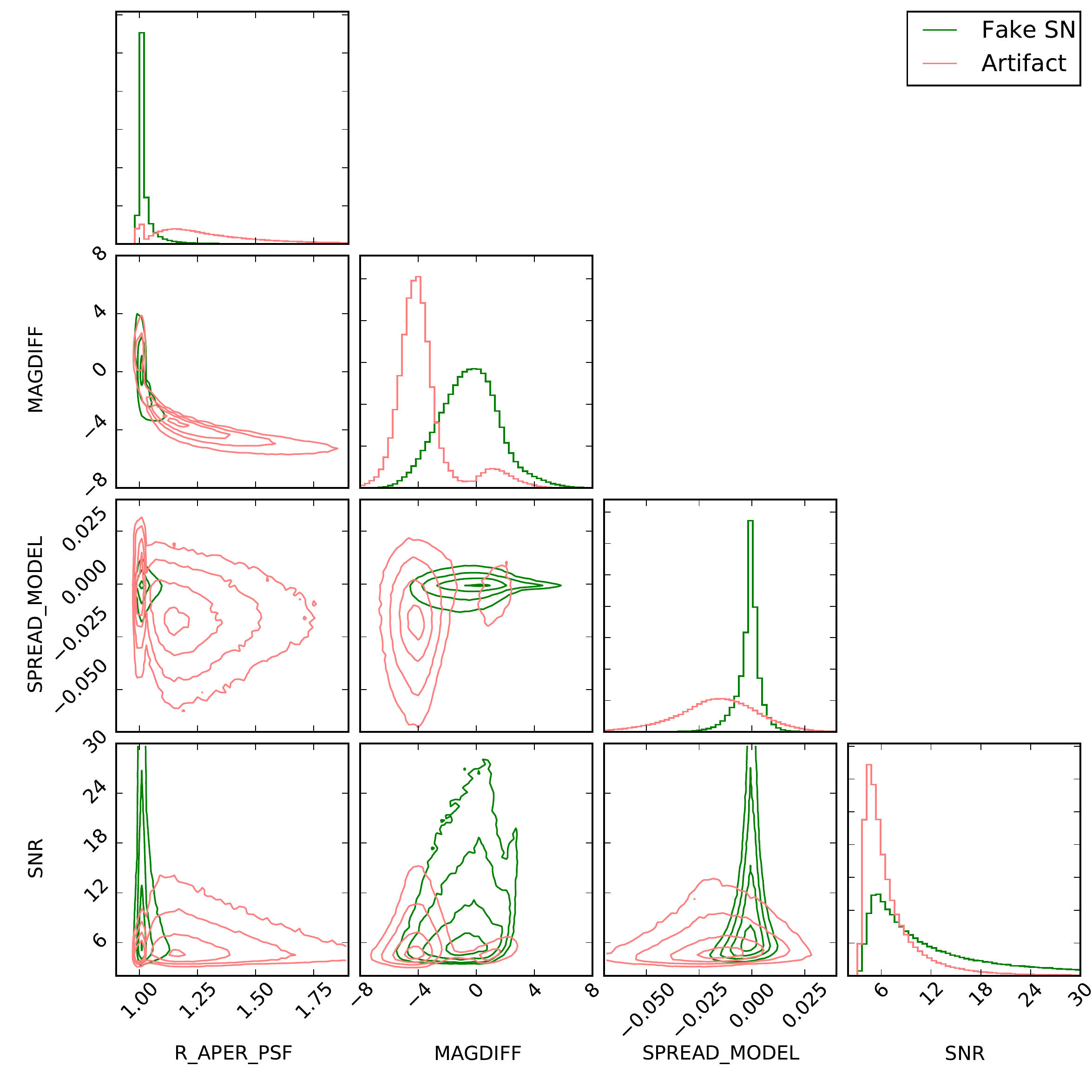}
 \caption{Contours of \q{r\_aper\_psf}, \q{magdiff},
 		  and \q{spread\_model}---the three most important
          features in the \aS~Random Forest model,
          computed
          using the feature importance evaluation scheme described in 
          \S \ref{section:importances}---and the signal-to-noise
          ratio, \q{snr}. 
          The importances of \q{r\_aper\_psf}, \q{magdiff},
          and \q{spread\_model} were 0.148, 0.094, and 0.066, respectively.
          The contours show that the relationships between the features
          are highly nonlinear and better suited to machine learning
          techniques than hard selection cuts.}
 \label{fig:top3features}
\end{figure*}

In this section we present new features developed for \aS.  
Let the superscripts $s, t,$ and $d$ on matrices defined in the
previous section denote
search, template, and difference images,
respectively.
The feature \q{r\_aper\_psf} is
designed to identify badly subtracted stars and galaxies 
on difference images
caused by poor astrometric alignment between search and
template images. These objects typically appear 
as overlapping circular regions 
of positive and negative flux colloquially known as
``dipoles." Examples are presented
in Figure \ref{fig:raperpsf}.
In these cases the typical search-template 
astrometric misalignment scale is comparable to 
the FWHM of the PSF, causing the contributions of
the negative and positive regions to
the total object-flux from a PSF-model fit
to be approximately equal in 
magnitude but opposite in sign, usually with a slight positive
excess as the PSF-fit is centered on the detection location,
where the flux is always positive. The total flux from
a PSF-model fit to a dipole is usually greater than but
comparable to the average flux per pixel in a five-pixel
circular aperture centered on the detection location on 
the template image. To this end, let $F_{aper, I}$ be the flux from a five-pixel circular aperture
centered on the location of a detection on the
uncompressed
template image. Let $F_{PSF, I}$ be the flux computed by fitting a
PSF-model to a $35 \times 35$ pixel cutout
centered on the location of the detection
on the uncompressed difference image. Then \q{r\_aper\_psf} is given by
\begin{equation}
        \q{r\_aper\_psf}=\frac{F_{aper, I} + F_{PSF,I}}{F_{PSF, I}}.
\end{equation}
We find that objects with \q{r\_aper\_psf} $> 1.25$ are almost
entirely ``dipoles.''

Let $a \in \{2, 3\}$, $b \in \{3, 5\}$. The four
features \q{n$a$sig$b$shift} represent the difference between the
number of pixels with flux values greater than or equal to $a$ in $(b +
2) \times (b + 2)$ element blocks centered on the detection position in
$R^{d}$ and $R^{t}$. These features coarsely describe changes in
the morphology of the source between the template and search images.

The feature \q{diffsum} is the sum of the matrix elements in a $5 \times 5$
element ($2.8 \times 2.8 \: \mathrm{arcsec}^2$) box centered on the
detection location in $R^{d}$. It is given by
\begin{equation}
\q{diffsum} = \sum_{i = -2}^{2} \sum_{j = -2}^{2} R^{d}_{x_c + i, y_c + j},
\end{equation}
where $x_c, y_c$ is the location of the central element on $R^d$. It gives a coarse measurement of the significance of the
detection.

\q{bandnum} is a numeric representation of the filter in which the
object was detected on the search image. This feature enables \aS~to
identify band-specific patterns.

\q{numneg} is intended to assess object-smoothness by
returning the number of negative elements in a $7\times7$ pixel
box centered on the object in $R^{d}$, exposing objects riddled with
negative pixels or objects that have a significant number of pixels
below $\med(R^{d})$. Used in concert with the S/N, \q{numneg} can help
identify high-S/N objects with spatial pixel intensity distributions
that do not vary smoothly, useful in rejecting hot pixels and cosmic
rays.

\q{lacosmic} was designed to identify cosmic rays and other objects
with spatial pixel intensity distributions that do not vary smoothly,
and is based loosely on the methodology that \cite{lacosmic} uses to
identify cosmic rays on arbitrary sky survey images. 
Derive the ``fine structure'' image $\mathbf{F}$ from
$\mathbf{B}^d$ according to 
\begin{equation}
 \mathbf{F} = (M_3 \ast \mathbf{B}^{d}) - ([M_3 \ast \mathbf{B}^{d}] \ast M_7),
\end{equation}
where $M_n$ is an $n \times n$ median filter. Then
\begin{equation}
 \q{lacosmic} = \max(\mathbf{B}^{d}) / \max(\mathbf{F}).
\end{equation}
Relatively speaking, this statistic should be large for objects that
do not vary smoothly, and small for objects that approximate a
PSF. The reader is referred to Figure 3 of \cite{lacosmic} for visual
examples. 

Bad columns and CCD edge effects that appear as
fuzzy vertical streaks near highly masked regions of
difference images are common types of artifacts. Because they share a number of visual
similarities, we designed a single feature, \q{colmeds}, to identify
them: 
\begin{equation}
\begin{split}
  \texttt{colmeds} = \max(\{\med(&\mathrm{transpose}(\mathbf{B}^{d})_i);\: \\
  &i \in \{0\ldots N_{col} - 1\}\}),
\end{split}                     
\end{equation}
where $N_{col}$ is the number of columns in $\mathbf{B}^d$. This
feature operates on the principle that the median of a column in
$\mathbf{B}^{d}$ should be comparable to the background if the cutout
is centered on a PSF, because, in general, even the column in which
the PSF is at its greatest spatial extent in $\mathbf{B}^{d}$ should still
contain more background pixels than source pixels. However, for
vertically oriented artifacts that occupy entire columns on
$\mathbf{B}^{d}$, this does not necessarily hold.
Since these artifacts frequently appear near masked regions
of images, we define \q{maskfrac} as the percentage of
$\mathbf{I}^{d}$ that is masked. 

The feature \q{spread\_model}
\citep{spreadmod1, spreadmod2} is a \q{SExtractor} star/galaxy separation
output parameter computed on the $\mathbf{I}^d$ cutout. It is a normalized simplified linear discriminant between the best fitting local PSF model and a slightly more extended model made from the same PSF convolved with a circular exponential disk model.

\subsubsection{Data Postprocessing}
\label{section:postprocessing}
When there is not a source in the galaxy coadd catalog within 5 arcsec
of an object detected on a difference image, certain classification
features cannot be computed for the object (see
Table \ref{tab:features}). If the feature of an object cannot be
computed, it is assigned the mean value of that feature from the
training set.

\subsection{Classification Algorithm Selection} 
\label{section:algorithms}
\begin{figure}
\centering \plotone{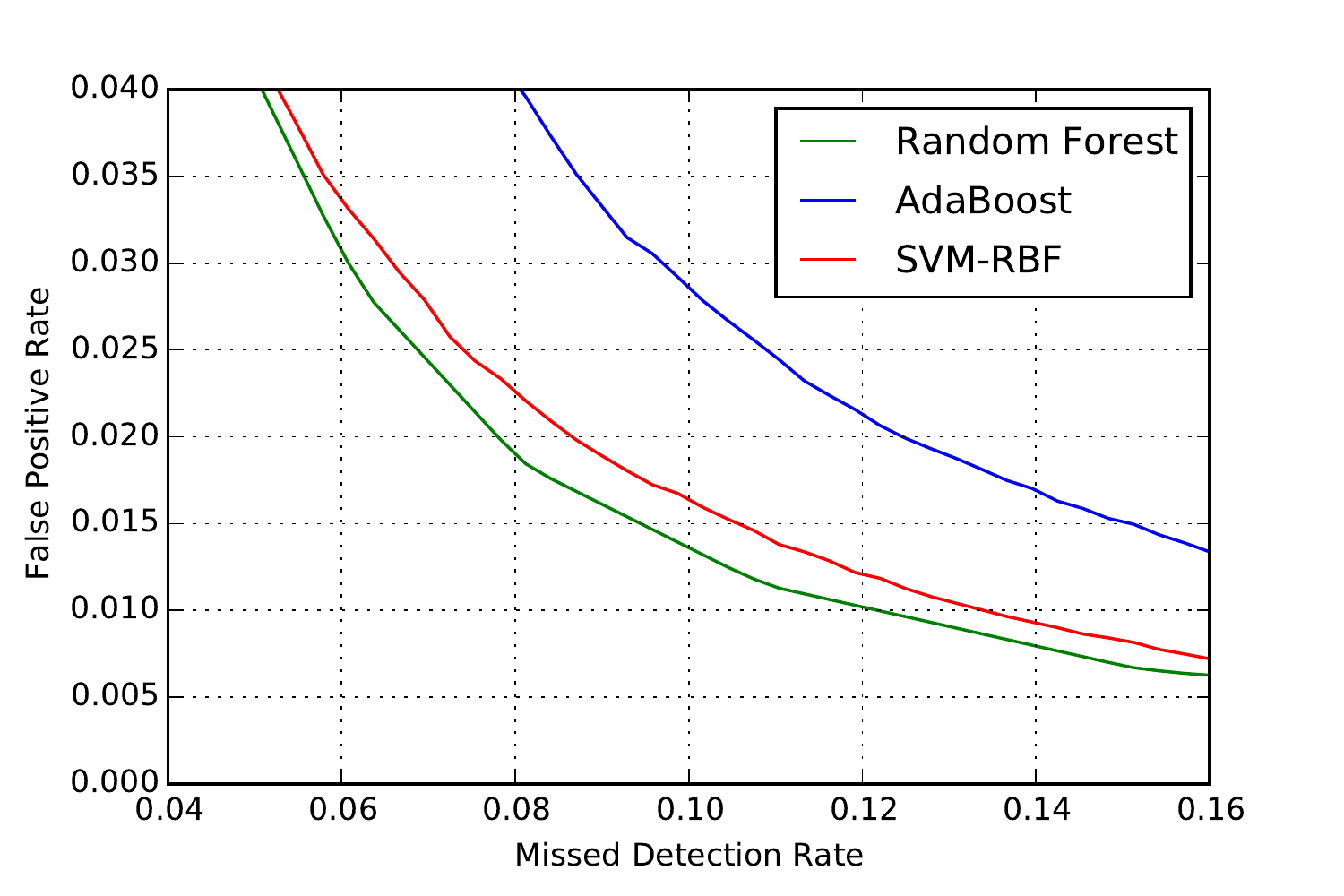}
\caption{Initial comparison of  the performance of a  Random Forest, a
  Support Vector Machine with a radial basis function kernel, and an
  AdaBoost Decision Tree classifier on the DES-SN artifact/non-artifact
  classification task.  Each classifier was trained on a randomly selected 67\%
  of the detections from a 100,000-detection subset of the training set, then tested on the
  remaining 33\%. This process was repeated three times until
  every detection in the subset was used in the testing set once.  The curves above
  represent the mean of each iteration.
  The closer a curve is
  to the origin, the better the 
  classifier. The
  unoptimized Random Forest outperformed the other two methods, and
  was selected.}
\label{fig:init}
\end{figure}

After we settled on an initial library of classification features, we
compared three well-known ML classification algorithms: a Random
Forest \citep{breiman}, a Support Vector Machine (SVM; \citealt{vapnik}), and an AdaBoost decision tree
classifier \citep{zhu}. We used \q{scikit-learn} \citep{sklearn}, an open source
Python package for machine learning, to instantiate examples of each
model with standard settings. We performed a three-fold cross-validated 
comparison using a randomly selected 100,000-detection subset of the training set described
in \S \ref{section:trainingdata}. 
The subset was used to 
avoid long training times for the SVM. For a description of cross validation
and the metrics used to evaluate each model,
see \S \ref{section:performance} and \S \ref{section:task}.  The
results appear in Figure \ref{fig:init}. We found that the performance
of all three models was comparable, but that the Random Forest
outperformed the other models by a small margin. We incorporated the
Random Forest model into \aS.

Random Forests are collections of
decision trees, or cascading
sequences of feature-space unit tests, that are constructed from
labeled training data.
For an introduction to decision trees, see \cite{ctrees}.
Random Forests can be used 
for predictive classification or regression. 
During the construction of a 
supervised Random Forest classifier, trees
in the forest are trained individually.
To construct a single tree, the training algorithm first chooses a bootstrapped sample of the
training data. The algorithm then attempts to recursively define a series of binary 
splits on the features of the training data
that optimally separate the training data into their constituent
classes.
During the construction of each node, a random subsample of features with a user-specified size is
selected with replacement. A fine grid of splits on each
feature is then defined, and the split that maximizes the increase in the purity of the incident training data is chosen for
the node. 
 
Two popular metrics for sample-purity  are the Gini coefficient \citep{gini} and the Shannon entropy \citep{shannon}. Define
the purity of a sample of difference image
objects to be\footnote{Some authors define
$P = \frac{\sum_{NA} w_{i}}{\sum_{NA} w_i + \sum_A w_i}$,
where $w_i$ is the weight of instance $i$, $\sum_A$ is
a sum over artifact events, and $\sum_{NA}$ is a sum
over non-artifact events. This
renders the definition of the Gini coefficient 
in Equation \ref{eqn:gini} as $\mathrm{Gini} = 
P(1-P)\sum_i w_i$.}
\begin{equation}
\label{eqn:purity}
P = \frac{N_{NA}}{N_A + N_{NA}},
\end{equation}
where $N_{NA}$ is the number of non-artifact
objects in the sample, and $N_A$ is the
number of artifacts in the sample. 
Note that $P=1$ for a sample composed
entirely of artifacts, $P=0$ for 
a sample composed entirely of 
non-artifacts, and $P(1-P) = 0$ 
for a sample composed entirely of
either artifacts or non-artifacts. Then the
Gini coefficient is
\begin{equation}
\label{eqn:gini}
\mathrm{Gini} = P(1-P)(N_A + N_{NA}).
\end{equation}
A tree with a Gini objective function
seeks at each node to minimize the quantity 
\begin{equation}
\mathrm{Gini}_\mathrm{lc} + \mathrm{Gini}_\mathrm{rc},
\end{equation}
where $\mathrm{Gini}_\mathrm{lc}$ is the Gini coefficient of the data incident on the node's
left child, and $\mathrm{Gini}_\mathrm{rc}$ is the
Gini coefficient of the data incident on the node's
right child. If $\mathrm{Gini}_\mathrm{lc} + \mathrm{Gini}_\mathrm{rc} > \mathrm{Gini}$, then
no split is performed and the node is declared
a terminal node. The process proceeds identically if another metric is used,
such as the Shannon entropy, the most
common alternative. The Shannon entropy
$S$ of a sample of difference image 
objects is given by
\begin{equation}
S = -p_{NA}\mathrm{log}_2(p_{NA}) - p_A\mathrm{log}_2(p_A),
\end{equation}
where $p_{NA}$ is the proportion of 
non-artifact objects in the sample,
and $p_A$ is the proportion of artifacts
in the sample. 

Nodes are generated in this fashion until a maximum depth or a user-specified measure of node purity is achieved. The number of
trees to grow in the forest is left as a free
parameter to be set by the
user. Training a single Random Forest using the
entire $\about 900,000$ object training sample with the hyperparameters selected from the 
grid search described in Table \ref{tab:grid}
took $\about 4.5$ minutes when the construction of the trees was distributed across 60 1.6GHz AMD Opteron 6262 HE processors.

Random Forests treat the classes of unseen
objects as unknown parameters that are
described probabilistically. 
An 
object to be 
classified descends each 
tree in the forest, beginning
at the root nodes.
Once a data point arrives at
a terminal node, the tree returns the fraction of the
training instances that reached that 
node that were labeled ``non-artifact."
The output of the trained \aS~Random Forest model on a single input data instance is the average of the outputs of each tree, representing the probability that the object 
is not an artifact, henceforth the 
``\aS~score" or ``ML score."
Ultimately,
a score of 0.5 was adopted as the cut $\tau$ to separate real detections
of astrophysical variability from artifacts in the DES-SN data; see \S\ref{section:cand} for
details. Class prediction for 200,000 unseen data instances took 9.5s on 
a single 1.6GHz AMD Opteron 6262 HE processor.

\subsection{Feature Importances}
\label{section:importances}
Numeric importances can be assigned to the 
features in a trained forest based on the 
amount of information they provided during
training \citep{ctrees}. For each tree $T$
in the forest, a tree-specific importance
for feature $i$ is computed according to 
\begin{equation}
\zeta_{i, T} = \sum_{n\in T} N(n)B\textcolor{blue}{_i}(n)\left[m(n) - m_{ch}(n)\right],
\end{equation}
where $n$ is an index over nodes in $T$, 
$N(n)$ is the number of 
training data points incident on node $n$, 
$B\textcolor{blue}{_i}(n)$ is $1$ if node $n$
splits on feature $i$ and $0$ otherwise, 
$m(n)$ is the value of the objective function
(usually
the Gini coefficient or the Shannon entropy,
see \S\ref{section:algorithms})
applied to the 
the training
data incident
on node $n$, and 
$m_{ch}(n)$ is the sum of the values
of the objective function applied to the
node's left and right children. 
The global importance of feature $i$
is the average of the tree-specific importances:
\begin{equation}
I_i = \frac{1}{N_T}\sum_T \zeta_{i, T},
\end{equation}
where $N_T$ is the number of trees in the
forest. 
In this article, importances
are normalized to sum to unity.

\subsection{Optimization} \label{section:optimization}
The construction of a Random Forest is governed by a number of free
parameters called hyperparameters.  The hyperparameters of
the Random Forest implementation used in this work
are \q{n\_estimators}, the number of decision trees in the forest,
\q{criterion},  the function  that  measures the  quality  of  a  proposed split
at a given tree node, \q{max\_features}, the number of features to
randomly select when looking for the best split at a given tree
node, \q{max\_depth}, the maximum depth of a
tree, and \q{min\_samples\_split}, the minimum number of samples required
to split an internal node.

\begin{deluxetable}{lr}
\tablewidth{0pt}
\tablecaption{Grid search results for \aS~hyperparameters. \label{tab:grid}}
\tablehead{\colhead{Hyperparameter} & \colhead{Values}}
\startdata
\texttt{n\_estimators} & 10, 50, \textbf{100},
300        \\        \texttt{criterion}        &        \texttt{gini},
\textbf{entropy}\\     \texttt{max\_features}     &     5, \textbf{6}\\  \texttt{min\_samples\_split}  &  2,  \textbf{3},  4,  10,  20,
50\\ \texttt{max\_depth} & \textbf{Unlimited}, 100, 30, 15, 5
\enddata
\tablecomments{A  3-fold  cross-validated search  over  the  grid of  Random
  Forest hyperparameters tabulated above was performed to characterize the
  the performance of the machine classifier. The hyperparameters of
  the best-performing classifier appear in bold.}
\end{deluxetable}

We performed  a 3-fold cross-validated (see \S \ref{section:task})  grid search over the  space of
Random Forest  hyperparameters described  in Table  \ref{tab:grid}.  A
total of 1,884 trainings were performed.  The best classifier
had 100 trees, used the Shannon entropy objective
function, chose 6 features for each split, required at least 3 samples
to split a node, and had unlimited depth, 
and it was incorporated into the code. Recursive feature elimination \citep{brink} was 
explored to improve the performance
of the classifier, but we found that
it provided no statistically significant
performance improvement.

\section{Performance}
\label{section:performance}
In this section, we describe
performance of \aS~on a realistic
classification task and the
effect of the code on the DES-SN transient candidate
scanning load.  
Performance statistics
for the classification task were measured using 
production Y1 data, whereas candidate-level 
effects were measured using
a complete reprocessing of Y1 data using an
updated difference imaging
pipeline. The reprocessed 
detection pool differed significantly 
from its production counterpart, providing
a out-of-sample data set for benchmarking the
effects of
the code on the scanning load.\footnote{
Although the re-processing of data
through the difference imaging pipeline from
the raw images is not useful for getting
spectra of live transients, it is quite
useful for acquiring host-galaxy targets 
for previously missed transients and 
is therefore performed regularly as
pipeline improvements are made.}

\subsection{Performance Metrics}
\label{section:metrics}

\begin{figure}
  \centering \includegraphics[width=64mm]{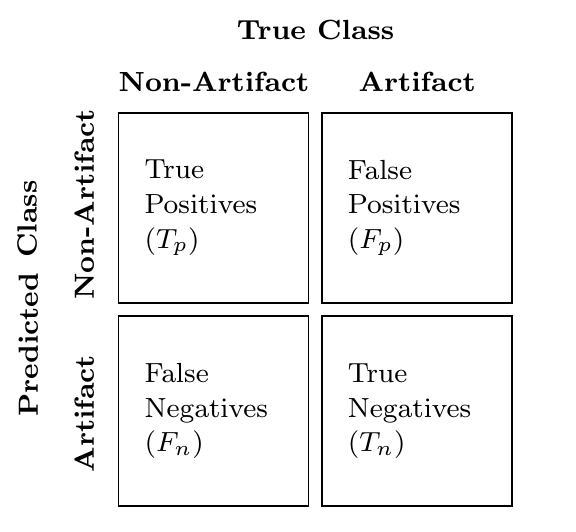} 
  \caption{Schematic
  confusion matrix for the \aS~classification task. Each matrix
  element $\mathbb{E}_{ij}$ represents the number of instances from
  the task's validation set with ground truth class label $j$ that
  were predicted to be members of class $i$.}  \label{fig:confmat}
\end{figure}

The performance of a classifier on an $n$-class task is completely
summarized by the corresponding $n\times n$ confusion matrix
$\mathbb{E}$, also known as a contingency table or error matrix.  The
matrix element $\mathbb{E}_{ij}$ represents the number of instances
from the task's validation set with ground truth class label $j$ that
were predicted to be members of class $i$.  A schematic $2 \times 2$
confusion matrix for the \aS~classification task is shown in
Figure \ref{fig:confmat}.

From the confusion matrix, several classifier performance metrics can
be computed. Two that frequently appear in the literature are the
False Positive Rate (FPR) and the Missed Detection Rate (MDR; also
known as the False Negative Rate or False Omission Rate). Using the
notation from Figure \ref{fig:confmat}, the FPR is defined by:
\begin{equation}
FPR = \frac{F_p}{F_p + T_n},
\end{equation}
and the missed detection rate by
\begin{equation}
MDR = \frac{F_n}{T_p + F_n}.
\end{equation}
For \aS, the FPR represents the fraction of artifacts in the
validation set that are predicted to be legitimate detections of
astrophysical variability. The MDR represents the fraction of
non-artifacts in the task's validation set that are
predicted to be artifacts. Another useful metric is the efficiency
or True Positive Rate (TPR),
\begin{equation}
\epsilon = \frac{T_p}{T_p + F_n},
\end{equation}
which represents the fraction of non-artifacts in the
sample that are classified correctly.
For the remainder of this study, we often refer to  the candidate-level
efficiency measured on fake SNe~Ia, $\epsilon_F$ (see \S\ref{section:cand}).  

Finally, the
receiver operating characteristic (ROC) is a graphical tool for
visualizing the performance of a classifier.  It displays FPR as a
function of MDR, both of which are parametric functions of $\tau$,
the \aS~score that one chooses to delineate the boundary between
``non-artifacts" and ``artifacts.''  One can use the ROC to
determine the location at which the trade-off between the FPR and MDR
is optimal for the survey at hand, a function of both the scanning
load and the potential bias introduced by the classifier, then solve
for the corresponding $\tau$.  By benchmarking the performance of the
classifier using the the ROC, one can paint a complete picture of its
performance that can also serve as a statistical guarantee on
performance in production, assuming a validation set and a production
data set that are identically distributed in feature space, and that
detections are scanned individually in production (see
\S \ref{section:cand}).

\subsection{Classification Task}
\label{section:task}

We used stratified 5-fold cross-validation to test the performance
of \aS.  Cross validation is a technique for assessing how the results
of a statistical analysis will generalize to an independent data set.
In a $k$-fold cross-validated analysis, a data set is partitioned into
$k$ disjoint subsets. $k$ iterations of training and testing are
performed. During the $i$th iteration, subset $i$ is held out as a
``validation'' set of labeled data instances that are not included in
the training sample, and the union of
the remaining $k-1$ subsets is passed to the classifier as a training
set. The classifier is trained and its predictive performance on the
validation set is recorded.  In standard $k$-fold cross-validation,
the partitioning of the original data set into disjoint subsets is
done by drawing samples at random without replacement from the
original data set. But in a stratified analysis, the drawing is
performed subject to the constraint that the distribution of classes
in each subset be the same as the distribution of classes in the
original data set. Cross-validation is useful because it enables one
to characterize how a classifier's performance varies with respect to
changes in the composition of training and testing data sets, helping
quantify and control ``generalization error.'' 

\subsection{Results}
\label{section:results}

\begin{figure}
        \centering
        \plotone{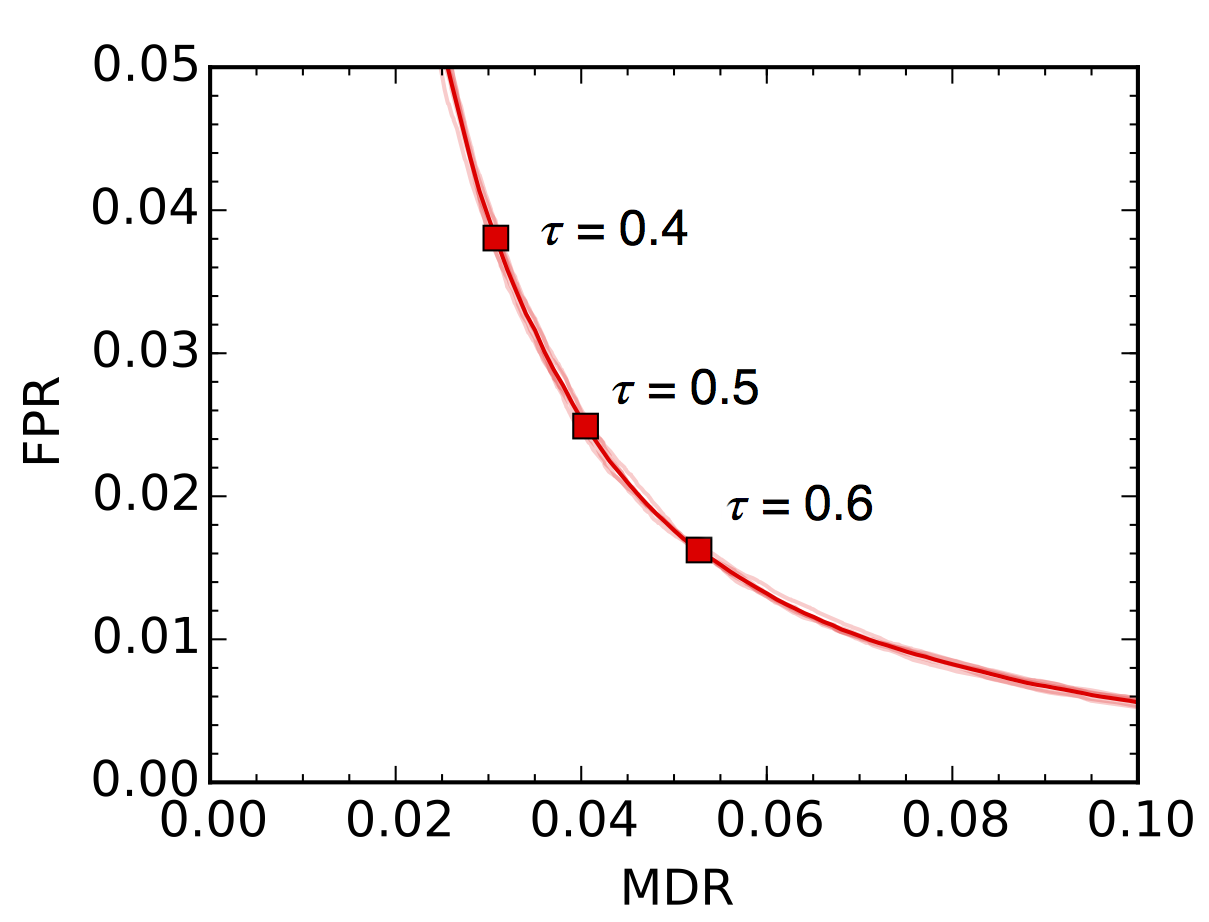}
        \caption{5-fold cross-validated 
        receiver operating characteristics
        of the best-performing classifier from
        \S\ref{section:optimization}. Six visually indistinguishable curves 
        are plotted: one translucent curve for each round
        of cross-validation, and one opaque curve 
        representing the mean. 
        Points on the mean ROC
        corresponding to different class discrimination boundaries
        $\tau$ are labeled. 
        $\tau = 0.5$ was adopted in DES-SN.}
        \label{fig:roc}
\end{figure}

\begin{figure*}
        \centering         
        \plotone{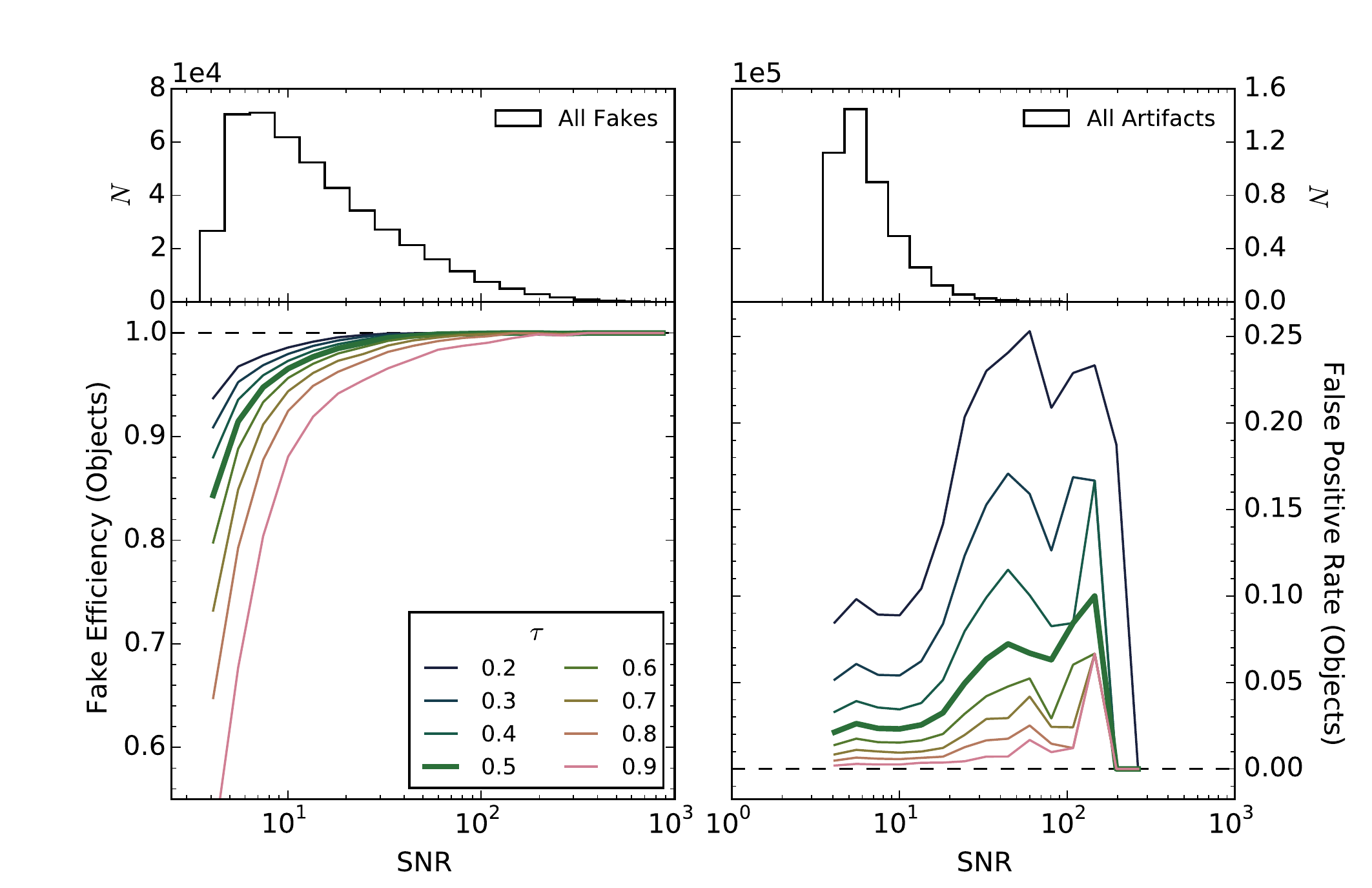}
        \caption{Object-level fake efficiency and false positive
        rate as a function of S/N, at several
                 \aS~score cuts. The S/N is computed by 
                 dividing the flux from a PSF-model fit to a $35 \times 35$ pixel cutout around the object in the difference image by the uncertainty from the fit. The artifact rejection efficiency
                 and missed detection rate are 1 minus the 
                 false positive rate and fake efficiency, respectively. The fake efficiency of \aS~degrades
                 at low S/N, whereas the false positive rate is
                 relatively constant in the S/N regime not dominated by small number statistics. $\tau = 0.5$ (bold) was adopted in DES-SN. }
        \label{fig:bias}
\end{figure*}

Figure \ref{fig:roc} shows the ROCs that resulted from each round of
cross-validation. We report that \aS~achieved an average
detection-level MDR of $4.0 \pm 0.1$ percent at a fixed FPR of 2.5
percent with $\tau = 0.5$, which was ultimately
adopted in the survey; see \S\ref{section:cand}.
We found that \aS~scores were correlated with detection signal-to-noise ratio
(S/N). Figure \ref{fig:bias} displays the fake efficiency and
false positive 
of \aS~using all out-of-sample detections of fake SNe
from each round of cross-validation.  At
$\mathrm{S/N} \lesssim 10$, the out-of-sample fake efficiency is
markedly lower than it is at higher S/N. The efficiency asymptotically
approaches unity for $\mathrm{S/N} \ga 100$. The effect becomes more
pronounced when the class discrimination boundary is raised. This
occurs because legitimate detections of astrophysical variability at
low S/N are similar to artifacts.
The false positive
rate remains relatively constant in the $\mathrm{S/N} \la 10$ regime, where the vast majority of artifacts reside. 

\subsection{Effect of \aS~on Transient Candidate Scanning Load}
\label{section:cand}

As discussed in \S\ref{section:des}, 
DES-SN performs target selection and scanning using
aggregates of spatially coincident detections from multiple nights and
filters (``candidates"). 
After the implementation of \aS,
the \q{NUMEPOCHS} requirement described in Table \ref{tab:cuts} was revised
to require that a candidate be detected on at least two distinct nights
having at least one detection with an ML
score greater than $\tau$ to become eligible for visual scanning. 
In this section we describe the effect of this revision on the
scanning load for an entire observing season using a full reprocessing of the Y1 data.

We sought to minimize
the size of our transient candidate scanning load with no more than a 1 percent
loss in $\epsilon_F$. By performing a grid search on
$\tau$, we found that we were able to reduce the number of candidates 
 during the first observing season of DES-SN  by a factor of 13.4, while maintaining
$\epsilon_F > 99.0$ per cent by
adopting $\tau = 0.5$. After implementing
\aS~using this $\tau$, we measured the quantity $\langle N_{A} / N_{NA} \rangle$,
 the average ratio of artifact objects to non-artifact detections that a
 human scanner encountered during a scanning session, using random
 samples of 3,000 objects drawn from the pool of objects passing the
 modified and unmodified cuts in Table \ref{tab:cuts}. We found that
 the ratio decreased by a factor of roughly 40 after the production
 implementation of \aS. Table \ref{tab:scanningload} summarizes these
 results.

\begin{deluxetable}{lccc}
\tablewidth{0pt}
\tablecaption{Effect of \aS~on Reprocessed DES Y1 Transient Candidate Scanning Load. \label{tab:scanningload}}
\tablehead{\colhead{} & \colhead{No ML} & \colhead{ML $(\tau=0.5 )$} & \colhead{ML / No ML}}
   \startdata 
   $N_{c}$\tablenotemark{a}  & 100,450 & 7,489 & 0.075\\
          $\langle N_A / N_{NA} \rangle$\tablenotemark{b}
                               & 13  & 0.34 & 0.027\\
          $\epsilon_{F}$\tablenotemark{c}
                               & 1.0 & 0.990  & 0.990
   \enddata
   \tablenotetext{a}{Total number of science candidates discovered.}
   \tablenotetext{b}{Average ratio of artifact to non-artifact detections in human scanning pool determined from scanning 3,000
   randomly selected detections from all science candidate detections.}
   					
   \tablenotetext{c}{\aS~candidate-level efficiency for fake SNe~Ia.}
\end{deluxetable}

\section{Discussion}
\label{section:discussion}
 \begin{figure*}
        \centering
        \includegraphics[width=160mm]{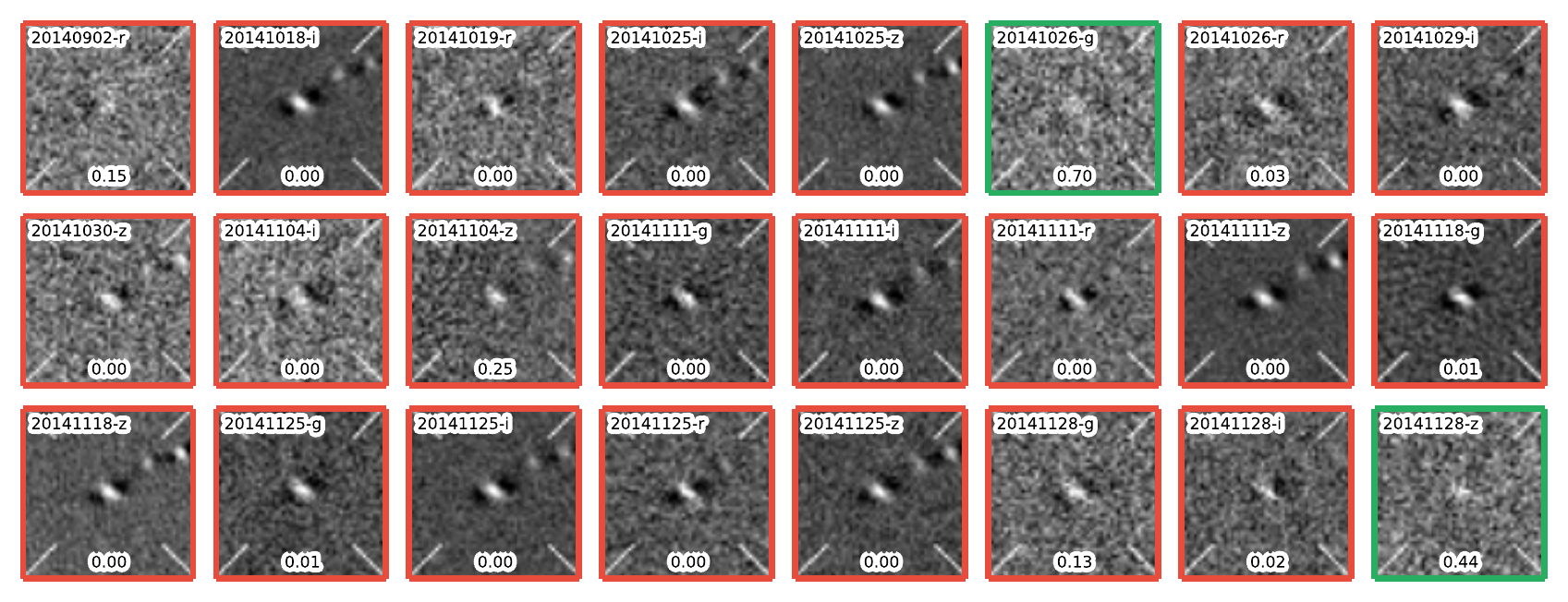}
        \caption{24 consecutively observed difference image cutouts of
          a poorly subtracted galaxy that was wrongly identified as a
          transient. The \aS~score of each detection appears at the
          bottom of each cutout. The mis-identification occurred because
          on two nights the candidate had a detection that received a
          score above an \aS~class discrimination boundary
          $\tau=0.4$ used during early code tests (green boxes).  
          Night-to-night variations in
          observing conditions, data reduction, and image subtraction
          can cause detections of artifacts to appear real. If a
          two-night trigger is used, spurious ``transients'' like this
          one can can easily accumulate as a season goes
          on. Consequently, care must be taken when using an artifact
          rejection framework that scores individual detections to
          make statements about aggregates of detections. Each image is
          labeled with the observation date and filter for the image,
          in the format YYYYMMDD-filter.}
       \label{fig:candidate_effect}
 \end{figure*}

With the development of \aS~and the use of fake overlays to robustly
measure efficiencies, the goal of automating artifact rejection on
difference images using supervised ML classification has reached a
certain level of maturity. With several historical and ongoing
time-domain surveys using ML techniques for candidate selection, it is
clear that the approach has been successful in improving astrophysical
source selection efficiency on images. However, there are still several
ways the process could be improved for large-scale transient searches
of the future, especially for ZTF and LSST, whose demands for
reliability, consistency, and transparency will eclipse those of
contemporary surveys. 

\subsection{Automating Artifact Rejection in Future Surveys}
\label{section:future}

For surveys like LSST and ZTF, small decreases in MDR are equivalent
to the recovery of vast numbers of new and interesting
transients. Decreasing the size of the feature set and increasing the
importance of each feature is one of the most direct routes to
decreasing MDR. However, designing and engineering effective
classification features is among the most time-consuming and least
intuitive aspects of framework design. Improving MDR by revising
feature sets is a matter of trial and error---occasionally,
performance improvements can result, but sometimes adding features can
degrade the performance of a classifier.  Ideally, surveys that will
retrain their classifiers periodically will have a rigorous,
deterministic procedure to extract the optimal feature set from a
given training data set. This is possible with the use of
convolutional neural networks (CNNs), a subclass of Artificial Neural
Networks, that can take images as input and infer an optimal set of
features for a given set of training data. The downside to CNNs is
that the resulting features are significantly more abstract than
astrophysically motivated features and consequently can be more
difficult to interpret, especially in comparison with Random Forests,
which assign each feature a relative importance. However, CNNs have
achieved high levels of performance for a diverse array of problems. They remain
relatively unexplored in the context of astrophysical data processing,
and bear examination for use in future surveys.

Next, unless great care is taken to produce a training data set that
is drawn from the same multidimensional feature distribution as the
testing data, dense regions of testing space might be completely
devoid of training data, leading to an unacceptable degradation of
classification accuracy in production. Developing a rigorous method
for avoiding such sample selection bias is crucial for future surveys,
for which small biases in the training set can result in meaningful
losses in efficiency.  The idea of incorporating active learning
techniques into astronomical ML classification frameworks has been
advanced as a technique for reducing sample selection bias \citep{active}.

Given a
testing set and a training set which are free to be drawn from
different distributions in feature space, in the pool-based active
learning for classification framework, an algorithm iteratively
selects, out of the entire set of unlabeled data, the object (or set
of objects) that would give the maximum performance gains for the
classification model, if its true label were known. The algorithm then
solicits a user to manually input the class of the object under
consideration, and then the object is automatically incorporated into
future training sets to improve upon the original classifier. Under
this paradigm, human scanners would play the valuable role of helping
the classifier learn from its mistakes, and each human hour spent
vetting data would immediately carry scientific return.  Active
learning could produce extremely powerful classifiers over short
timescales when used in concert with generative models for training
data. Instead of relying on historical data to train artifact
rejection algorithms during commissioning phases, experiments like
LSST could use generative models for survey observations to simulate
new data sets. After training a classifier using simulated data, in
production active learning could be used to automatically fill in gaps
in classifier knowledge and augment predictive accuracy.

In this work, we used a generative model of SN~Ia observations---overlaying fake SNe~Ia onto real host galaxies---to
produce the ``Non-Artifact'' component of our training data set. However, the
nearly 500,000 artifacts in our training set were human-scanned,
implying that future surveys will still need to do a great deal of
scanning before being able to get an ML classifier off the ground.  A
new survey should not intentionally alter the pipeline to produce
artifacts during commissioning, as it is crucial that the unseen data
be drawn from the same feature distributions as the training data.
For surveys with $\langle N_A / N_{NA} \rangle \gtrsim 100$,
\cite{brink} showed that a robust artifact library can be prepared by
randomly sampling from all detections of variability produced by the difference imaging pipeline. For surveys or pipelines
that do not produce as many artifacts, some initial scanning to
produce a few $10^4$-artifact library from commissioning data should
be sufficient to produce an initial training set \citep{brink, pca}.

\subsection{Eliminating Spurious Candidates}
\label{section:discusscand}

Using a two-night trigger, some spurious science candidates can be
created due to nightly
variations in astrometry, observing conditions, and repeatedly imaged
source brightnesses that cause night-to-night fluctuations in the
appearance of candidates on difference images. These variations lead
to a spread of ML scores for a given candidate.  
As an observing season progress, artifacts can accumulate
large numbers of detections via repeated visits. Although
for a typical artifact the vast majority of detections
fail the ML requirement, the fluctuations in ML scores
can cause a small fraction of the detections to satisfy
the \aS~requirement. Figure \ref{fig:candidate_effect} shows an example of this effect.  

Mitigating the buildup of spurious multi-night candidates
could be achieved by implementing a second ML
classification framework that takes as input multi-night
information, including the detection-level output of
\aS, to predict whether a given science candidate represents 
a bona-fide astrophysical source. Training
data compilation could be performed by randomly 
selecting time-contiguous strings of detections from 
known candidates. The lengths of the strings could be drawn from
a distribution specified during framework development. 
Candidate-level features could characterize the temporal variation
of detection level features, such as the highest and lowest night-to-night
shifts in 
\aS~score, magnitude, and astrometric uncertainty.

\acknowledgments
DAG thanks an anonymous referee
for comments that improved
the paper. 
We are grateful for the extraordinary contributions of our CTIO
  colleagues and the DES Camera, Commissioning and Science
  Verification teams for achieving excellent instrument and telescope
  conditions that have made this work possible.  The success of this
  project also relies critically on the expertise and dedication of
  the DES Data Management organization. Funding for DES projects has
  been provided by the U.S. Department of Energy, the U.S. National
  Science Foundation, the Ministry of Science and Education of Spain,
  the Science and Technology Facilities Council of the United Kingdom,
  the Higher Education Funding Council for England, the National
  Center for Supercomputing Applications at the University of Illinois
  at Urbana-Champaign, the Kavli Institute of Cosmological Physics at
  the University of Chicago, Financiadora de Estudos e Projetos,
  Funda\c{c}\~{a}o Carlos Chagas Filho de Amparo \'{a} Pesquisa do
  Estado do Rio de Janeiro, Conselho Nacional de Desenvolvimento
  Cient\'{i}fico e Tecnol\'{o}gico and the Minist\'{e}rio da
  Ci\^{e}ncia e Tecnologia, the Deutsche Forschungsgemeinschaft and
  the collaborating institutions in the Dark Energy Survey.

  The collaborating institutions are Argonne National Laboratory, the
  University of California, Santa Cruz, the University of Cambridge,
  Centro de Investigaciones Energeticas, Medioambientales y
  Tecnologicas-Madrid, the University of Chicago, University College
  London, the DES-Brazil Consortium, the Eidgen\"{o}ssische Tecnische
  Hochschule (ETH) Z\"{u}rich, Fermi National Accelerator Laboratory,
  the University of Edinburgh, the University of Illinois at
  Urbana-Champaign, the Institut de Ciencies de l'Espai (IEEC/CSIC),
  the Intitut de Fisica d'Altes Energies, Lawrence Berkeley National
  Laboratory, the Ludwig-Maximilians Universit\"{a}t and the
  associated Excellence Cluster Universe, the University of Michigan,
  the National Optical Astronomy Observatory, the University of
  Nottingham, the Ohio State University, the University of
  Pennsylvania, the University of Portsmouth, SLAC National Acclerator
  Laboratory, Stanford University, the University of Sussex, and Texas
  A\&M University.

  This research used resources of the National Energy Research
  Scientific Computing Center, a DOE Office of Science User Facility
  supported by the Office of Science of the U.S. Department of Energy
  under Contract No. DE-AC02-05CH11231. Figure \ref{fig:top3features}
  was generated with a modified version of \q{triangle.py} \citep{tpy}. 
  ACR acknowledges financial support provided by the 
  PAPDRJ CAPES/FAPERJ Fellowship.
  FS acknowledges financial support provided by CAPES under contract No. 3171-13-2. The DES participants from Spanish institutions are partially supported by MINECO under grants AYA2012-39559, ESP2013-48274, FPA2013-47986, and Centro de Excelencia Severo Ochoa SEV-2012-0234, some of which include ERDF funds from the European Union.

\end{document}